\documentclass[12pt]{iopart}
\usepackage{iopams}
\usepackage{braket}
\usepackage{graphicx}
\usepackage{subfigure}
\usepackage{cite}
\usepackage{algorithm}
\usepackage{algorithmic}

\begin{document}
\newcommand{\swave}[0]{$\it{s}$-wave}
\newcommand{\pwave}[0]{$\it{p}$-wave}
\newcommand{\K}{$^{40}\rm{K}$}
\newcommand{\Rb}{$^{87}\rm{Rb}$}
\newcommand{\us}{$\rm{\mu s}$}
\newcommand{\mT}{$\rm{mT}$}
\newcommand{\ez}{$\bi{e}_z$}
\newcommand{\ex}{$\bi{e}_x$}
\newcommand{\um}{$\rm{\mu m}$}
\renewcommand{\thefootnote}{\arabic{footnote}}

\title[]{Direct observation of Feshbach enhanced \swave{} scattering of fermions}

\author{D Genkina$^{1}$\footnotemark[1], LM Aycock$^{1,2}$\footnotemark[1], BK Stuhl$^1$, H-I Lu$^1$ RA Williams$^3$, and IB Spielman$^1$\footnote{Corresponding author}}
\footnotetext[1]{These authors contributed equally to the work}

\address{$^1$Joint Quantum Institute, National Institute of Standards and Technology, and University of Maryland, Gaithersburg, MD, 20899 USA}
\address{$^2$Physics Department, Cornell University, Ithaca, NY 14850 USA}
\address{$^3$National Physical Laboratory, Hampton Road, Teddington, Middlesex, TW11 0LW, UK}

\ead{ian.spielman@nist.gov}

\begin{abstract}
We directly measured the normalized \swave{} scattering cross-section of ultracold \K{} atoms across a magnetic-field Feshbach resonance by colliding pairs of degenerate Fermi gases (DFGs) and imaging the scattered atoms. We extracted the scattered fraction for a range of bias magnetic fields, and measured the resonance location to be $B_0 = 20.206(15)$  \mT{} with width $\Delta = 1.0(5)$  \mT{}. To optimize the signal-to-noise ratio of atom number in scattering images, we developed techniques to interpret absorption images in a regime where recoil induced detuning corrections are significant. These imaging techniques are generally applicable to experiments with lighter alkalis that would benefit from maximizing signal-to-noise ratio on atom number counting at the expense of spatial imaging resolution.
\end{abstract}

\vspace{2pc}

\maketitle

\section{Introduction}
Feshbach resonances are widely used for tuning the interaction strength in ultracold atomic gases. In degenerate Fermi gases (DFGs), the tunability of interactions provided by Feshbach resonances has allowed for studies of the creation of molecular Bose-Einstein Condensates (BECs) \cite{Greiner03,Zwierlein03, Jochim03} as well as observation of the phase transition from the Bardeen-Cooper-Schrieffer (BCS) superconducting regime to the BEC regime at sufficiently low temperatures \cite{Bartenstein04, Bourdel04, Zwierlein04, Regal04}. Conversely, measuring interactions as a function of controlled parameters can be used to characterize a Feshbach resonance.

\par A Feshbach resonance occurs when a diatomic molecular state energetically approaches the two-atom continuum \cite{Chin10, Timmermans99}.  For a magnetic-field Feshbach resonance, a bias magnetic field defines the relative energy of the free atomic states in two hyperfine sublevels and the molecular state. Consequently, the Feshbach resonance can be accessed by changing the bias field. In cold atomic systems where only \swave{} channels contribute to scattering, the interactions are entirely characterized by the scattering length $a$. In the simple case where there are no inelastic two-body channels, such as for the \K{} resonance discussed in this work, the effect of the resonance on the scattering length between two free atoms is \cite{Chin10}
\begin{equation}
a(B)=a_{\rm{bg}}\left(1-\frac{\Delta}{B-B_0}\right),
\label{feshbachEq}
\end{equation}
where $a_{\rm{bg}}$ is the scattering length far from any resonance (background scattering length), $\Delta$ is the width of the resonance, and $B_0$ is the field value at which the resonance occurs.

The exact value of the resonant field $B_0$ is difficult to calculate analytically and is commonly computed via numerical models based on experimental input parameters \cite{Tiesinga93, Lysebo09, Gao11} or determined experimentally \cite{Inouye98, Cornish00}. Many experimental techniques have been used to characterize Feshbach resonances, including the observation of atom loss due to three-body inelastic scattering, measurement of re-thermalization timescales, and anisotropic expansion of a cloud upon release from a confining potential, all of which infer the elastic scattering cross section from collective behavior of the cloud \cite{Regal03,OHara02,Monroe93}. 

Here direct scattering was the primary probe of the location and width of a Feshbach resonance. We collided pairs of DFGs and imaged the resulting \swave{} scattered atoms as a function of bias magnetic field. This allowed us to observe the enhancement in scattering without relying on proxy effects. We measured the fraction of atoms scattered during the collision at different bias magnetic fields and deduced the location and width of the resonance.

In contrast to BECs, where scattering halos are readily imaged \cite{Chikkatur00,Wilson04,Williams2012}, the density of Fermi clouds is typically $\approx$ 100 times less that that of BECs \footnote{This is not the case for recently realized erbium and dysprosium DFGs \cite{Aikawa14,Lu12}, where strong dipolar interactions are present}, making it necessary to enhance the strength of inter-atomic interactions to directly detect the scattered atoms. In our dilute DFGs, even with the resonant enhancement of the scattering cross section, only a small fraction of the atoms scattered. Using typical absorption imaging, direct detection of scattered atoms was difficult due to detection uncertainty that particularly affected regions of low atomic density. To optimize the signal-to-noise ratio (SNR) for low atom numbers, we absorption imaged with fairly long, high-intensity pulses \--- a non-standard regime \---  which imparted a non-negligible velocity and therefore Doppler shift to the atoms.  Simulation of the absorption imaging process was necessary for an accurate interpretation of these images. Using the simulation-corrected images, we extracted the fraction of atoms scattered in our collision experiment.

This paper is divided into two parts. First, we study absorption imaging in the presence of a significant time-dependent Doppler shift and show how we use our results to interpret data. Second, we describe our \swave{} scattering experiment and extract a measure of the location and width of the Feshbach resonance in \K{}.

\section{Absorption imaging in the presence of strong recoil induced detuning}
\label{sec:2}
Absorption imaging measures the shadow cast by an atomic ensemble in an illuminating probe laser beam with angular frequency $\omega_{\rm{L}}$. This imaging technique relies on optical transitions between ground and excited atomic states. Such atomic transitions have an energy difference $\hbar\omega_0$, and a natural transition linewidth $\Gamma$. When interacting with a laser field an atom scatters photons from the field into the vacuum modes. In the two-level atom approximation, the rate of scattering is \cite{LCT}
\begin{equation}
\gamma_{\rm{sc}}= \frac{\Gamma}{2} \frac{\tilde{I}}{1+4\tilde{\delta}^2 +\tilde{I}},
\label{eq:scatrate}
\end{equation}
where $\tilde{I}=I/I_{\rm{sat}}$ is the laser intensity in units of the saturation intensity, and $\tilde{\delta}=\delta/\Gamma$ is the detuning $\delta=\omega_{\rm{L}}-\omega_0$ in units of the natural linewidth. 

\begin{figure}
	\subfigure[]{\includegraphics*[scale=0.5]{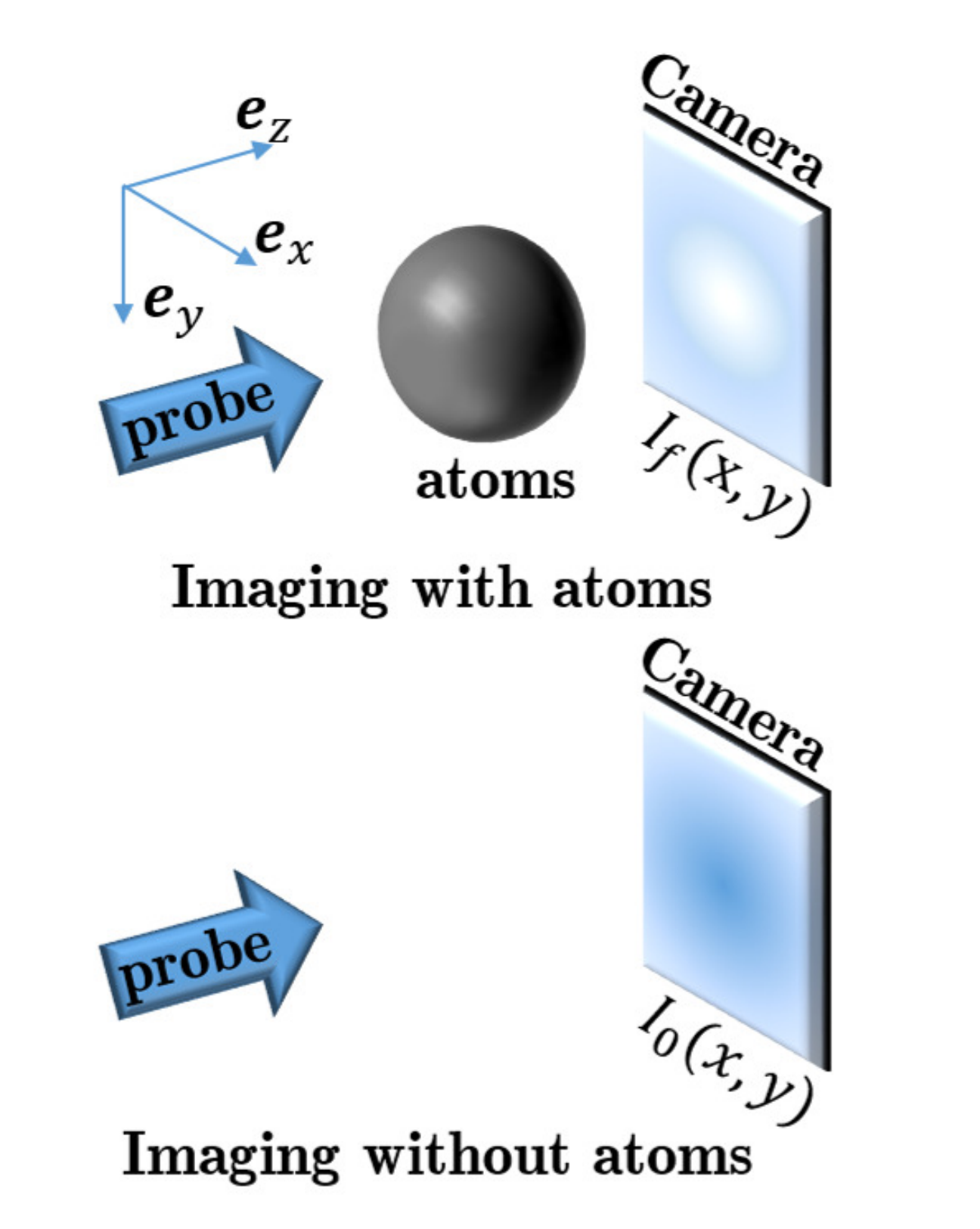}}
	\subfigure[]{\includegraphics*{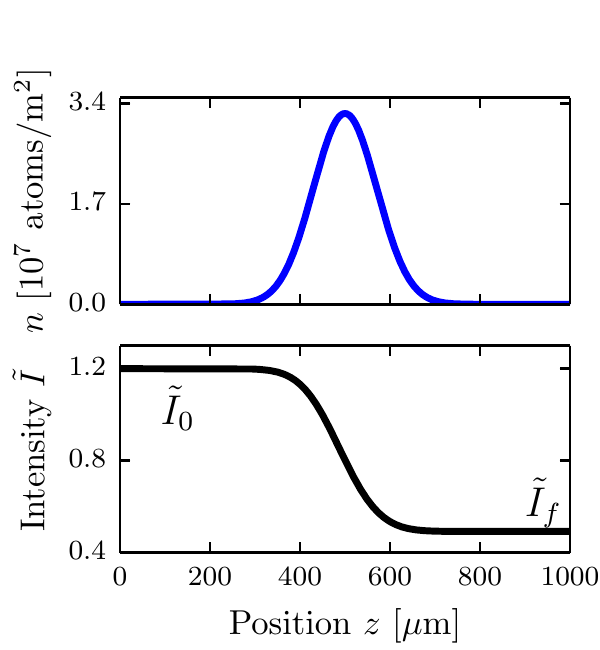}}
	\subfigure[]{\includegraphics*[scale=0.55]{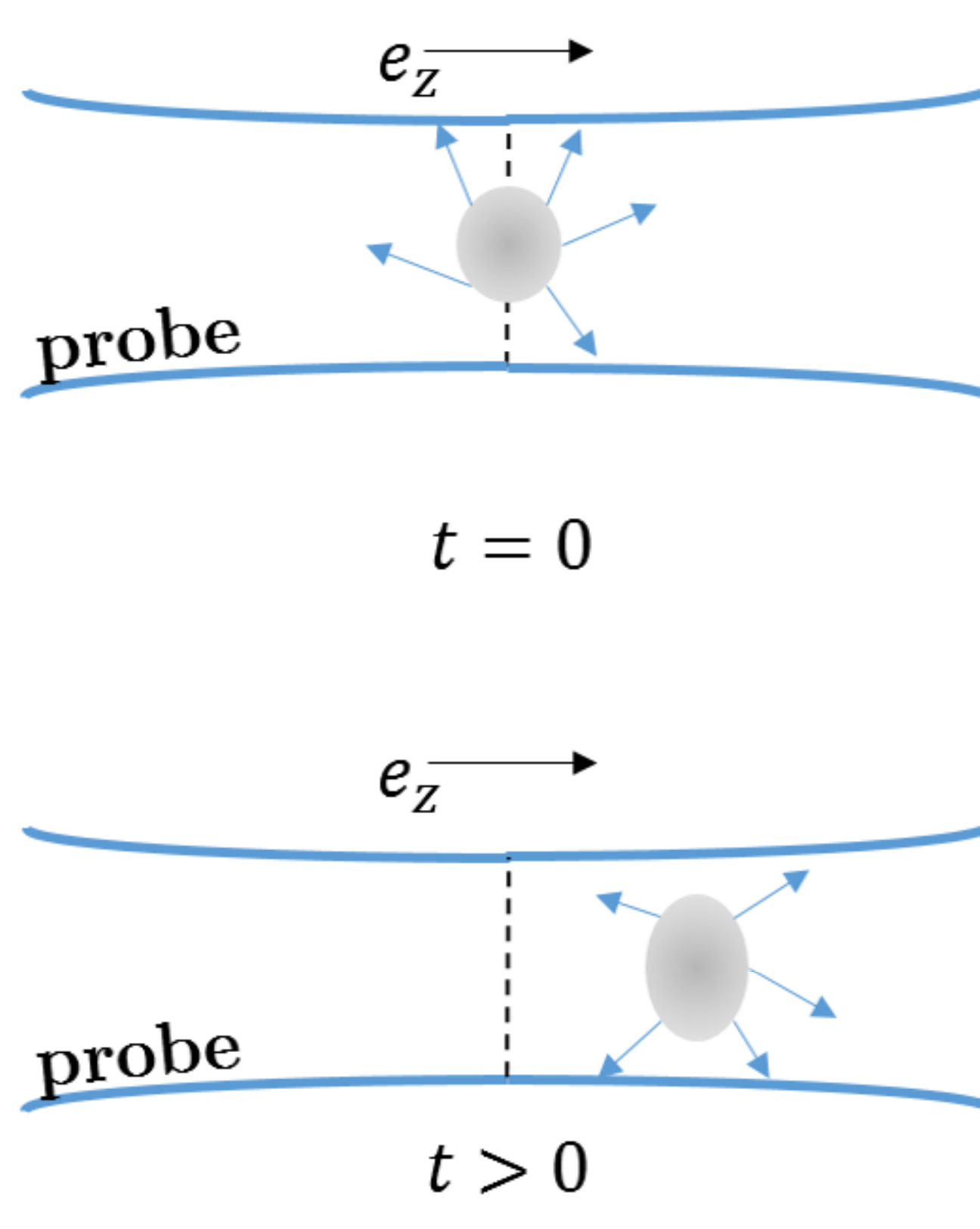}}
\caption{Absorption imaging. (a) Near resonant probe light illuminates the atoms, and the transmitted light (containing a shadow of the atoms) is imaged on the camera. A second image taken with no atoms provides a reference. (b)  The probe beam is partially absorbed as it traverses the cloud, and the intensity seen by atoms further along the imaging direction \ez{} is lowered.  (c) An atomic cloud illuminated by a probe light field absorbs photons from the probe and re-emits them in all directions. This process results in a net acceleration of the cloud in the direction of the probe light as well as diffusive spreading in the transverse directions.  }
\label{fig:absorptionIntor}
\end{figure}

An absorption image is obtained by shining an on- or near-resonant probe beam (generally $\tilde{\delta}\ll1$) onto the atomic cloud. Some of the light is scattered by the atoms, and the shadow cast by the atoms in the probe beam, $\tilde{I}_f(x,y)$, is imaged onto a camera, as depicted in Fig. \ref{fig:absorptionIntor}(a) (top). The probe light is reapplied with the atoms absent to calibrate the intensity $\tilde{I}_0(x,y)$ of light unaffected by the atoms (bottom).

Consider the light as it travels along the imaging axis \ez{} through a 3D atomic density profile $\rho(x,y,z)$. We focus on a single pixel of the camera: sensitive to a single column of atoms $\rho(z)$, integrated in x and y over the pixel, giving a single value of $\tilde{I}_0$ and $\tilde{I}_f$. Every atom scatters light according to Eq. (\ref{eq:scatrate}). Therefore, the atoms further along the imaging axis \ez{} experience a reduced optical intensity due to attenuation of the laser field by the other atoms (Fig. \ref{fig:absorptionIntor}(b)). The intensity change from scattering as a function of $z$ is
\begin{equation}
\frac{d\tilde{I}(z)}{dz}=-\hbar\omega_{\rm{L}}\rho(z)\gamma_{sc}(z)=-\rho(z)\sigma_0\frac{\tilde{I}(z)}{1+4\tilde{\delta}^2 +\tilde{I}},
\label{eq:dIdz}
\end{equation}
where $\sigma_0$ is the resonant scattering cross section. Integrating this equation  \cite{Reinaudi07} yields a straightforward relation between the observed intensities $\tilde{I}_0$ and $\tilde{I}_f$ and the atomic column density $n = \int \rho\left(z\right) \,\mathrm{d}z$:
\begin{equation}
\sigma_0 n = -\left(1+4\tilde{\delta}^2\right)\ln\left(I_f/I_0\right)+ \tilde{I}_0-\tilde{I}_f.
\label{eq2}
\end{equation}
We call the column density deduced from this relation $\sigma_0 n^{\rm{(1)}}$. When the probe intensity is much smaller than the saturation intensity, $\tilde{I}_0\ll1$, and the probe light is on resonance, $\tilde{\delta}=0$, the right hand side of Eq. (\ref{eq2}) reduces to the optical depth, defined as $OD=-\ln \left(I_f/I_0\right)$ \cite{Reinaudi07}, giving the simple relationship $\sigma_0 n^{\rm{(0)}}=OD$. In all other regimes, the optical depth is not constant and depends on the probe intensity and imaging time.

Equations (\ref{eq:dIdz})-(\ref{eq2}) neglect the atomic recoil momentum and the resulting Doppler shift \cite{Konstantinidis12}. When an atom absorbs a photon from the laser light field it acquires a momentum kick $\hbar  k_{\rm{r}}$ in the \ez{} direction. The associated recoil velocity is $v_{\rm{r}}=\hbar k_{\rm{r}}/m$, where $m$ is the atomic mass and $\hbar k_{\rm{r}}= h/\lambda$ is the recoil momentum from the laser with wavelength $\lambda$. Each re-emitted photon imparts a similar recoil momentum $\bi{p}_e$, but over many scattering events this momentum distribution averages to zero. Therefore the atom  will only acquire an average velocity per photon $v_{\rm{r}}$ along \ez{}. The variance of $\bi{p}_e$, however, is not zero, allowing the atoms to acquire some momentum transverse to the laser field. While we ignore this correction, it results in the reduction of spatial resolution in the final image and its effect on the atomic cloud is pictured in Fig. \ref{fig:absorptionIntor}(c).

\begin{figure}
	\includegraphics{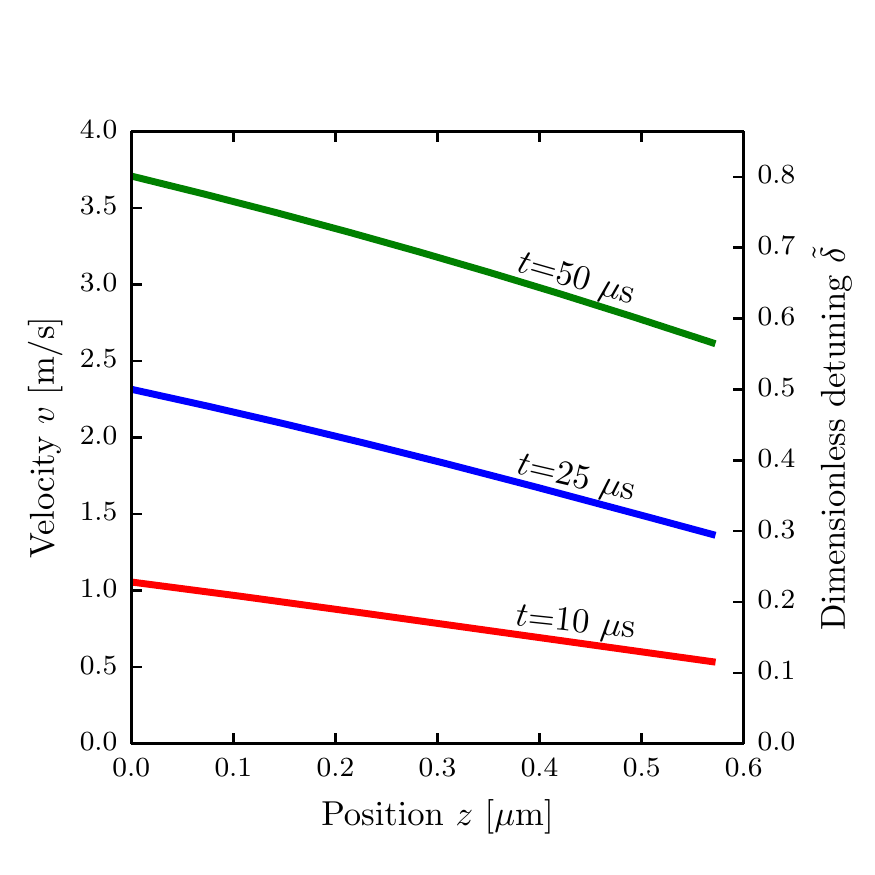}
\caption{Dependence of velocity and detuning on position simulated for \K{} at three different imaging times and a probe intensity $\tilde{I}_0=0.8$.}
\label{fig:expos}
\end{figure}

The average atomic velocity parallel to the light field after scattering $N$ photons is $N v_{\rm{r}}$ and the laser frequency as seen by the atoms is Doppler shifted  $\delta= k_{\rm{r}} N v_{\rm{r}}$ from resonance. After an atom scatters $N_{\rm{photons}}=\Gamma/2 k_{\rm{r}}v_{\rm{r}}$, it gets Doppler detuned by half a linewidth. For a probe intensity of $\tilde{I}_0=1$, the time it takes a single atom to scatter on average that many photons is given by $t_{\rm{recoil}}=N_{\rm{photons}}/\gamma_{\rm{sc}}=2/ k_{\rm{r}}v_{\rm{r}}$. For \K{} atoms imaged on the D2 transition, the case relevant for our experiment, $N_{\rm{photons}}=178$ and $t_{\rm{recoil}}=18.76$ \us{} \--- for imaging times longer than that the recoil induced detuning correction cannot be neglected. Furthermore, this detuning varies both with imaging time $t$ and with distance along the propagation direction \ez{} (Fig. \ref{fig:expos}). Thus, the laser's spatially varying intensity profile in the atomic cloud also depends on time:
\begin{equation}
\frac{d\tilde{I}(t,z)}{dz}=-\sigma_0 \rho(t,z) \frac{\tilde{I}(t,z)}{1+4\tilde{\delta}(t,z)^2 +\tilde{I}(t,z)}. \label{eq3}
\end{equation}
Assuming that the atoms do not move significantly during the imaging time (we will remove this assumption shortly), the dimensionless detuning is
\begin{equation}
\tilde{\delta}(t,z)=\frac{ k_{\rm{r}} v_{\rm{r}}}{2\sigma_0 \rho(t,z)}\int_0^t \frac{d\tilde{I}(z,\tau)}{dz}\,\mathrm{d}\tau; \label{eq4}
\end{equation}
the relationship between the atomic density and the observed intensities is no longer straightforward. Peturbative treatments of these equations also prove insufficient (see \ref{AppendixA1}).

\subsection{Simulation}
\label{simulSec}
To obtain a relationship between between the atomic density and the observed intensities in this non-standard regime, we numerically simulated the imaging process, including the recoil induced detuning. We used parameters relevant to our experiment \--- the D2 transition of \K{}, with $\lambda$ = 766.701 nm, $\Gamma/2 \pi$ = 6.035 MHz, $v_r$ = 1.302 cm/s and $I_{\rm{sat}}$ = 1.75 mW/cm$^2$ \cite{Tiecke}. The simulation obtained $\tilde{I}_f$ as a function of imaging time $t$, atomic density $\sigma_0 n$, and probe intensity $\tilde{I}_0$. 

We performed two versions of this simulation. First, we took a simplistic approach where the spatial distribution of atoms did not change appreciably during the imaging time: $vt\ll I_0/(\hbar \omega_{\rm{L}} \gamma_{\rm{sc}} \rho)$ \--- the stationary assumption. Starting with a Gaussian density profile, we numerically integrated Eqs. (\ref{eq3}) and (\ref{eq4}) and obtained a simulated optical depth for a range of input probe intensities and atomic column densities.  We used the results of this simulation to check the self-consistency of the  stationary atom assumption and found it to be invalid (see \ref{AppendixA2}).

To account for the changing atomic distribution during the imaging pulse, we numerically simulated the classical kinetics of atoms subject to the recoil driven optical forces, and obtained a dynamics adjusted version of the simulated optical depth. We compared the optical depths predicted by each of the two simulations in the parameter range  $0.3\leq t \leq 100$ \us{}, $0.01\leq\tilde{I}_0 \leq 50$ and $0.01\leq \sigma_0 n \leq 2$ and found that the predicted optical depths were hardly changed by including the full time evolution (see \ref{AppendixA3}).  Thus, for the purposes of deducing the atomic density from experimental optical depths, the stationary atom simulation is sufficient in the experimentally relevant parameter regime we explored. Furthermore, we simulated a range of initial density profiles $\rho(z)$, and found their impact on the simulated $OD$ to be negligible \--- the only observable is the integrated atomic density $n$, and 3D atomic densities cannot be reconstructed. 

\begin{figure}
	\includegraphics{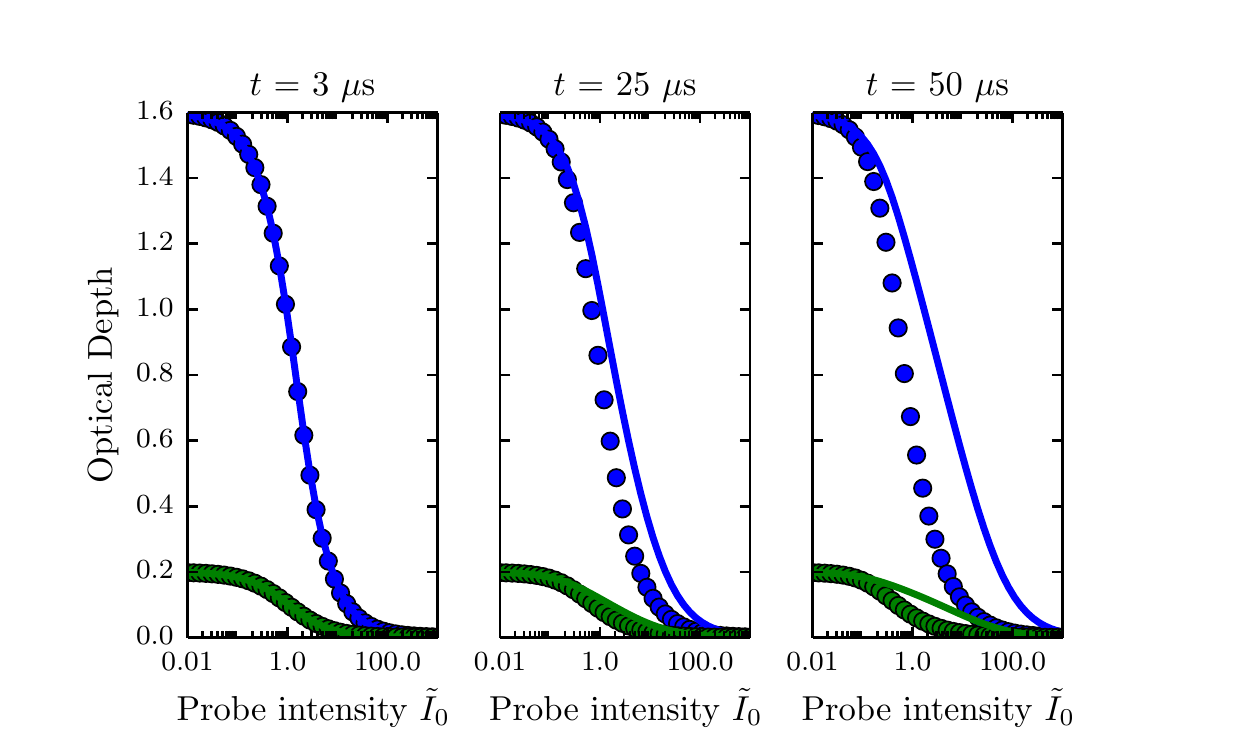}
\caption{Optical depth as a function of probe intensity as predicted by the simulation (dots) and by Eq. (\ref{eq2}) (curves), for three different imaging times, and for column densities $\sigma_0 n=$ 1.6 (blue) and 0.2 (green). As expected, the predictions agree in both the high and low intensity limits, and differ for probe intensities comparable to the saturation intensity and longer imaging times. }
\label{fig:IsatLimits}
\end{figure}

 Figure \ref{fig:IsatLimits} illustrates the effect of the recoil induced detuning correction as obtained from our simulations. In the limit of low probe intensity, $\tilde{I}_0\ll 1$, the atomic velocities are hardly changed and the recoil induced detuning correction is negligible. In the limit of high probe intensity  $\tilde{I}_0\gg \tilde{\delta}$, even far detuned atoms will scatter light at their maximum rate and the overall absorption will again be unaffected by the correction. In the intermediate regime, there is a significant deviation between the optical depth predicted by Eq. (\ref{eq2}) and the simulated optical depth, and this deviation becomes stronger with longer imaging times.  

This simulation provided us with a correction procedure to interpret experimentally observed $\tilde{I}_f$ and $\tilde{I}_0$. For a given imaging time, the simulation predicted a final intensity as a function of probe intensity and atomic column density. We inverted this to prediction to obtain an atomic column density given our observed intensities. For interpreting experimental images, we used the optical depths predicted by the traveling atom simulation, $OD^{\rm{sim}}$.

\subsection{SNR optimization}
\label{optimSec}
We added shot noise to our simulation and established optimal imaging parameters to maximize the SNR of this detection scheme. We considered Poisson noise on the detected arriving photons (i.e., photoelectrons) with standard deviation proportional to $\sqrt{q_{\rm{e}} N_p}$, where $q_{\rm{e}}$ is the quantum efficiency of the camera (0.66 for our camera) and $N_p$ is the photon number. We then propagated this uncertainty through the correction scheme described in Sec. \ref{simulSec} to obtain the uncertainty in a deduced column density, $ \delta_{\sigma_0 n}$. We define the SNR as $\sigma_0 n/\delta_{\sigma_0 n}$.

\begin{figure}
	\subfigure[]{\includegraphics{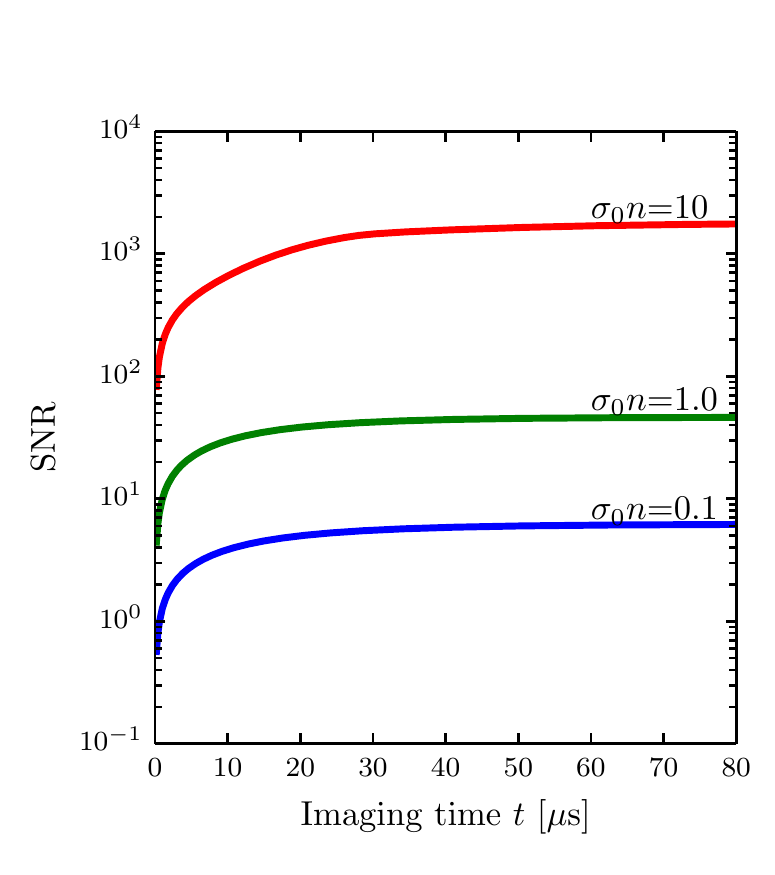}}
	\subfigure[]{\includegraphics{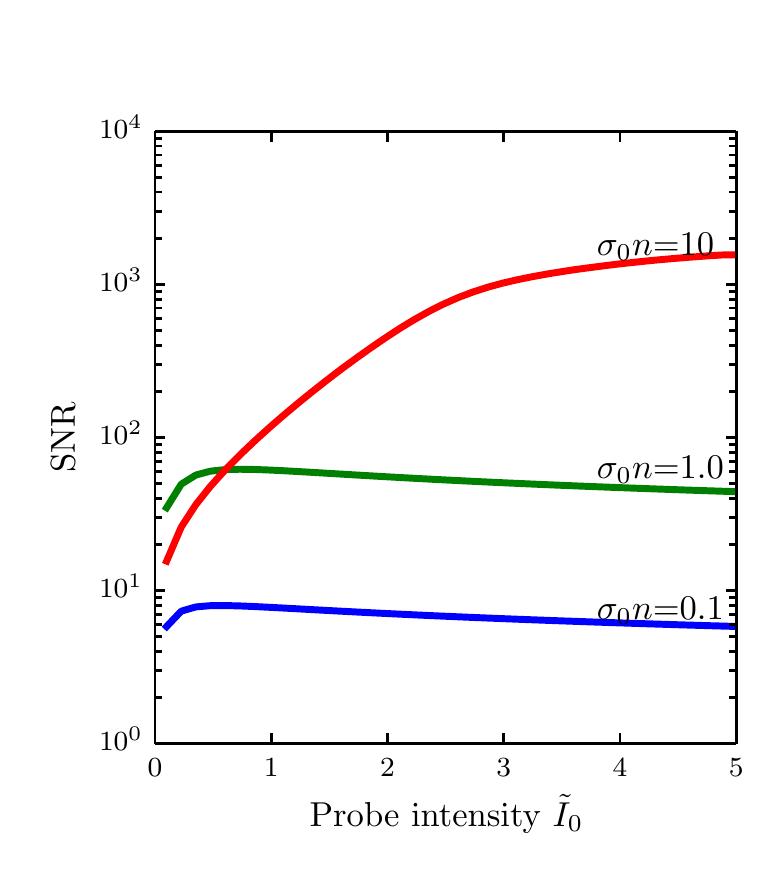}}
\caption{SNR for three different column densities after correcting for recoil induced detuning. (a) SNR as a function of imaging time for a probe intensity of $\tilde{I}_0=5.0$ and (b) SNR as a function of probe intensity for an imaging time of 50 \us{}.}
\label{fig:SNR}
\end{figure}

As seen in Fig. \ref{fig:SNR}(a), after about 40 \us{} extending the imaging time no longer yields appreciable improvement in SNR. Imaging for 40 \us{} as opposed to 10 \us{}, where the uncorrected model is appropriate, improves the SNR by a factor of  1.5. We therefore performed the experiments described in the second section at 40 \us{} imaging time. Figure \ref{fig:SNR}(b) shows that the optimal probe intensity varies with the atomic column density. For low atom numbers, $\sigma_0 n\approx0.1$, a probe intensity of $\tilde{I}_0\approx0.6$ is best. However, in our experiment the probe intensity had a Gaussian profile and was not uniform over the whole image.  The typical probe intensities used in our experiments varied over the $\tilde{I}_0=0.1-0.7$  range.

\subsection{Calibration of saturation intensity}
The calibration of the observed signal in units of the saturation intensity is crucial to our measurement of the column densities. Our absorption images were taken using a charge-coupled device (CCD) camera. For each pixel, the camera returned an integer number of counts proportional to the radiant fluence seen by that pixel.  However, the proportionality constant depended on many factors, such as the quantum efficiency of the camera, the electronic gain during the readout process, losses in the imaging system. One way to determine this proportionality constant is to experimentally calibrate the saturation intensity in counts per unit time.

\begin{figure}
	\includegraphics{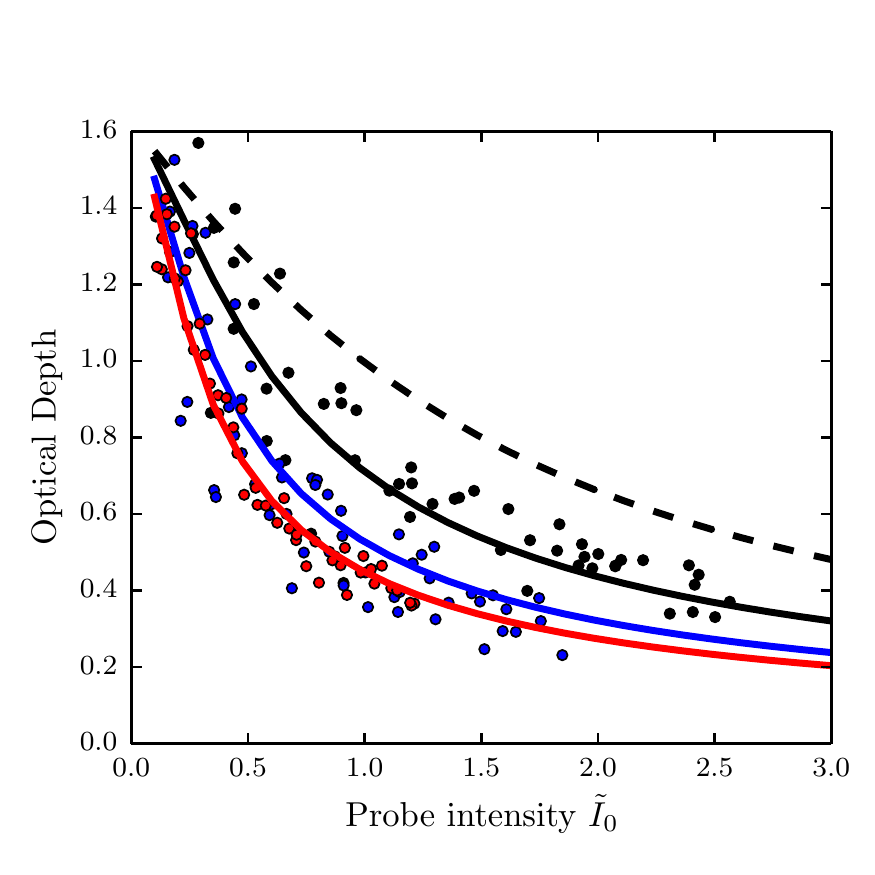}
\caption{The optical depth as a function of probe intensity for three imaging times: $t=40$ \us{} (black),  $t=75$ \us{} (blue),  $t=100$ \us{} (red). The dots represent experimental data and the curves represent the best fit of simulated data. The optimal fit parameters pictured are a $\sigma_0 n$ of 1.627(5) and saturation intensity of 29(7) counts/\us{}. The dashed curve represents the theoretical prediction without recoil induced detuning corrections.}
\label{fig:isatCalib}
\end{figure}

To calibrate the saturation intensity in camera counts per unit time, we took absorption images of  the atoms at three different imaging times (40 \us{}, 100 \us{}, and 200 \us{}) with varying probe intensities. For each image we obtained $\tilde{I}_0$ and $\tilde{I}_f$ in counts per microsecond by averaging over a few pixels in a region of constant atomic column density. We then simultaneously fit our simulated optical depth  $OD^{\rm{sim}}$ to this full data set, with the atomic density $\sigma_0 n$ and  $I_{\rm{sat}}$ in counts per microsecond as free parameters. As seen in Fig. \ref{fig:isatCalib}, the model produced a good fit to the experimental data, and provided a calibration of the saturation intensity for our experiment.

\section{\swave{} scattering experiment}
For our measurement of the Feshbach resonance location and width, we collided two counter-propagating \K{} clouds in a spin mixture of $\ket{F=9/2,m_F=-9/2}$ and $\ket{F=9/2,m_F=-7/2}$ hyperfine states and observed the resulting \swave{} halo of scattered atoms.  We measured the dependence of the scattered atomic fraction on the bias magnetic field in the vicinity of the Feshbach resonance. We used this data to extract the location of the resonance at 20.206(15) \mT{} with width 1.0(5) \mT{}, consistent with the accepted values of 20.210(7) \mT{} and 0.78(6) \mT{} \cite{Regal04}.
\subsection{Experimental procedure}
Our experiment is a hybrid \K{} and \Rb{} apparatus, previously described in \cite{Williams13, Lin09, KarinaThesis}. Initially, we prepared a spin polarized  $\ket{F=9/2,m_F=-9/2}$  DFG of $\approx 4\times10^5$ atoms of \K{} at a temperature of  $T\approx0.4\, T_{\rm{F}}$, where $ T_{\rm{F}}$ is the Fermi temperature, in a crossed optical dipole trap with frequencies $(\omega_x,\omega_y,\omega_z)/2\pi =(39, 42, 124)$ Hz (see \ref{AppendixB}). 

To map out the entire Feshbach resonance without the added losses associated with going through the resonance \cite{Chin10}, we needed to create equal spin mixtures  of $\ket{F=9/2,m_F=-9/2}$ and $\ket{F=9/2,m_F=-7/2}$ on either side of the resonance. We ramped the bias magnetic field to 19.05 mT (21.71 mT) and turned on a 42.42 MHz (47.11 MHz) rf field resonant with the Zeeman splitting between the two states when preparing the mixture below (above) the Feshbach resonance. We then  sinusoidally modulated the bias field at 125 Hz for 0.5 s, with a 0.14 mT amplitude, producing an equal mixture of the two hyperfine states. The depolarization allowed the fermions to re-thermalize, allowing us to further evaporate in the dipole trap \cite{DeMarco99}. These hyperfine states of \K{} were then used to study their Feshbach resonance.

After evaporation, we ramped the bias field in a two-step fashion to the desired field value near the Feshbach resonance. The two-step procedure was designed to allow us to approach the set-point quickly and avoid additional losses. This procedure used two sets of Helmholtz coils \---  large coils that provided the majority of the bias field but had a long inductive timescale, and smaller coils only capable of generating 0.59 mT of bias, but with a shorter inductive timescale. We approached the field using the large coils to bring the magnetic field to a set-point 0.59 \mT{} above or below the intended bias field. We held the atoms at this field for 100 $\rm{ms}$ to allow the eddy currents induced by the large coils to settle, and then used the smaller coils to quickly change the bias field the remaining 0.59 \mT{}.  For all set-points, the data was taken approaching from both above and below the Feshbach resonance \footnote{An extra data point was taken on each side far from the resonance using only one approach.}.

Once at the intended bias field, we split the cloud into two spatially overlapping components with opposite momenta  and observed scattering as they separated. These counterpropagating components were created using Kapitza-Dirac pulses of a 1D retro-reflected near-resonant optical lattice ($\lambda_{\rm{L}}$=766.704 nm) with 8$E_{\rm{L}}$ depth, where $E_{\rm{L}}=\hbar^2 k_{\rm{L}}^2/2m_{\rm{K}}$ is the lattice recoil energy and $\hbar k_{\rm{L}}=2\pi \hbar/ \lambda$ is the recoil momentum. We rapidly pulsed this lattice on and off with a double-pulse protocol \cite{Wu05}. The pulse sequence was optimized to transfer most of the atoms into the $\pm 2 \hbar k_{\rm{L}}$ momentum states. Since the initial Fermi gas had a wide momentum spread (here, $ 2 \hbar k_{\rm{L}}\approx2.5\hbar k_{\rm{F}}$, where $k_{\rm{F}}$ is the Fermi momentum), and the lattice pulsing is a momentum dependent process \cite{Edwards10}, not all the atoms were transferred into the target momentum states. We optimized our pulse times to minimize the atoms remaining in the zero momentum state. 

We then released the atoms from the trap and allowed 1 ms for the two opposite momentum states to pass through each other while interacting at the magnetic field set-point. For data taken approaching the set-point from below, we then ramped down the field and imaged the atoms. For data taken approaching the set-point from above, molecules may have been created when crossing the Feshbach resonance. Therefore, we first ramped the field up to a point above the resonance to dissociate any molecules that were created and then quickly ramped the field back down and imaged the atoms. After a total time-of-flight $t_{TOF}=6.8$ $\rm{ms}$, we used a 40 \us{} imaging pulse with $\tilde{I}_0 \approx 0.6$ at the center of the probe laser, chosen for SNR optimization as described in Sec. \ref{optimSec}.

\subsection{Magnetic field calibration}

The magnetic fields produced by our coils in the regime of interest were independently calibrated by rf-spectroscopy on the $\ket{F=9/2, m_F=-9/2}$ to $\ket{F=9/2, m_F=-7/2}$ transition. We prepared a spin polarized state and ramped the large coils to variable set-points. We then illuminated the atoms with a rf field with frequency $\nu_{rf}$ and performed adiabatic rapid passage (ARP) by ramping the smaller coils 0.0284 mT in 250 ms. We applied a Stern-Gerlach pulse and imaged the atoms to measure the fractional population in the $\ket{F=9/2, m_F=-9/2}$  and $\ket{F=9/2, m_F=-7/2}$ states. We fit the fractional population as a function of current to a Gaussian function\footnote{Due to our low rf coupling and high noise, we did not fit to the traditional Loretzian model}. The center of the Gaussian corresponded to the resonant magnetic field, which was produced by the high inductance coil setpoint plus half the ARP, 0.0142 mT, with an uncertainty given by the Gaussian width. We used the Breit-Rabi formula to determine the resonant field value at $\nu_{rf}$. We did this for 5 different rf frequencies, and acquired a field calibration with an uncertainty of 0.004 mT, which was included in the listed uncertainty on our measured value of $B_0$.

\subsection{Methods}

\begin{figure}
	\includegraphics{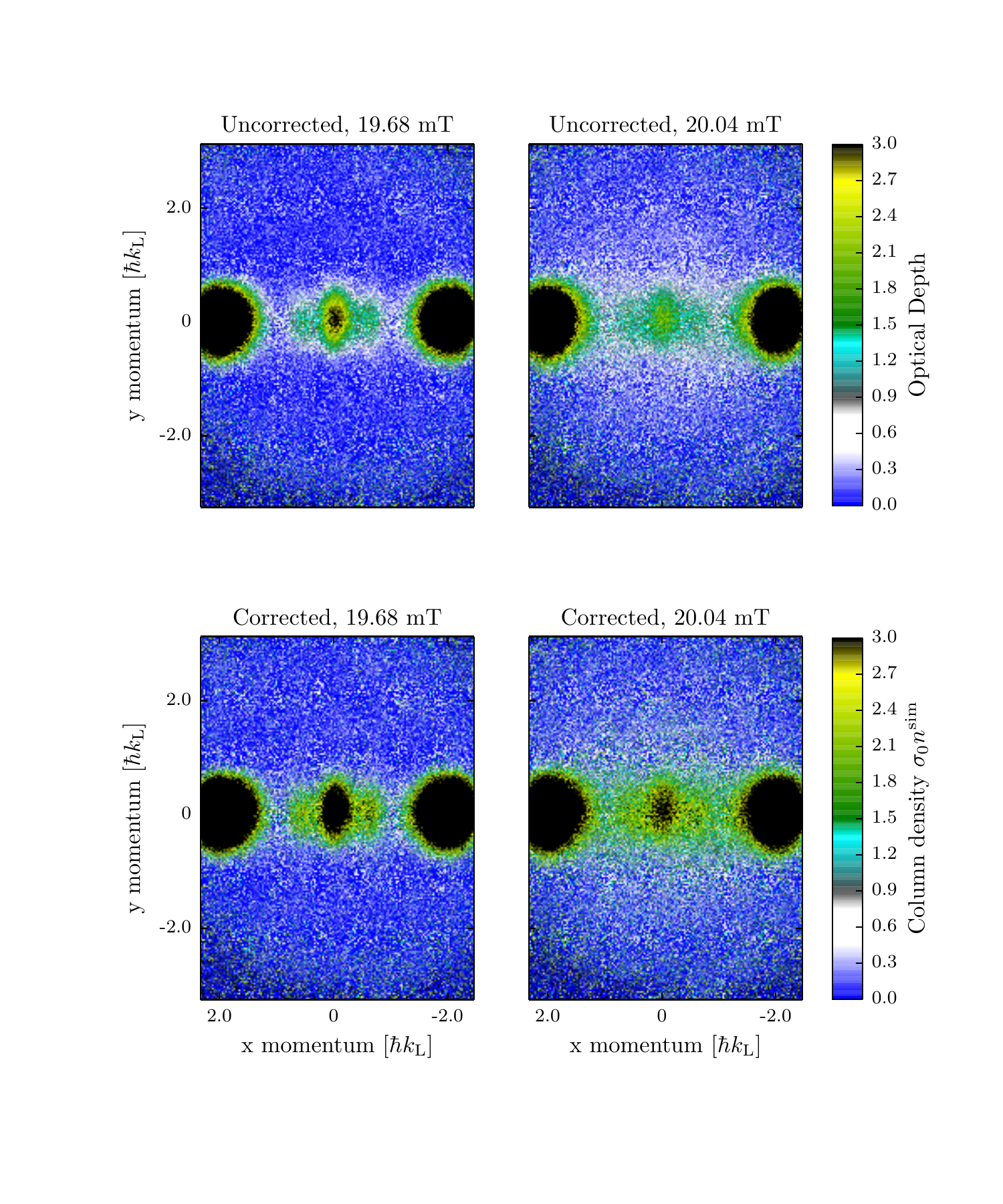}
\caption{Examples of our absorption images after 6.8 ms TOF. The 1D lattice imparts momentum along \ex{}. In each image, the two large clouds on the left and right are the atoms in the $\pm 2 k_{\rm{L}}$ momentum orders that passed through each other unscattered. The smaller cloud in the center is composed of the atoms that remained in the lowest band of the lattice after pulsing, and thus obtained no momentum. The thin spread of atoms around these clouds are the atoms that underwent scattering.   (Top) Raw optical depths,  far from resonance (19.68 mT) on the left and close to resonance (20.04 mT) on the right, (Bottom) atomic column density $\sigma_0 n^{\rm{sim}}$ obtained by applying corrections to raw optical depth above based on simulations (Sec. \ref{simulSec}).}
\label{fig:SampleCorrection}
\end{figure}

We first processed the \swave{} scattering images by comparing the observed $OD$ to simulations taking into account the recoil induced detuning as described in Sec. \ref{sec:2}. An example of images before and after processing are shown in Fig. \ref{fig:SampleCorrection}. The processing constituted a $\approx 30\,\%$ change in the column density.   

\begin{figure}
	\subfigure[]{\includegraphics[scale=0.23]{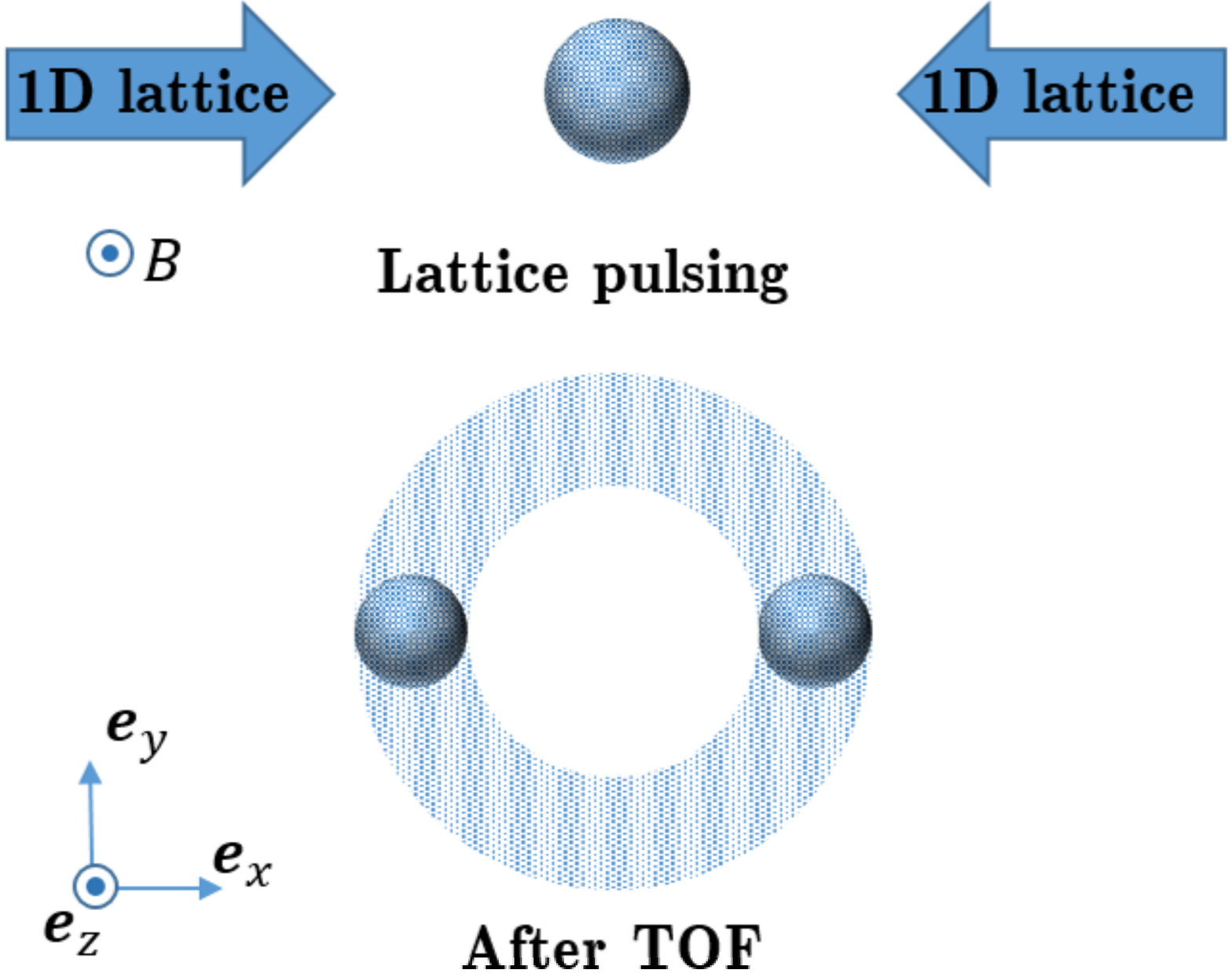}}
	\subfigure[]{\includegraphics{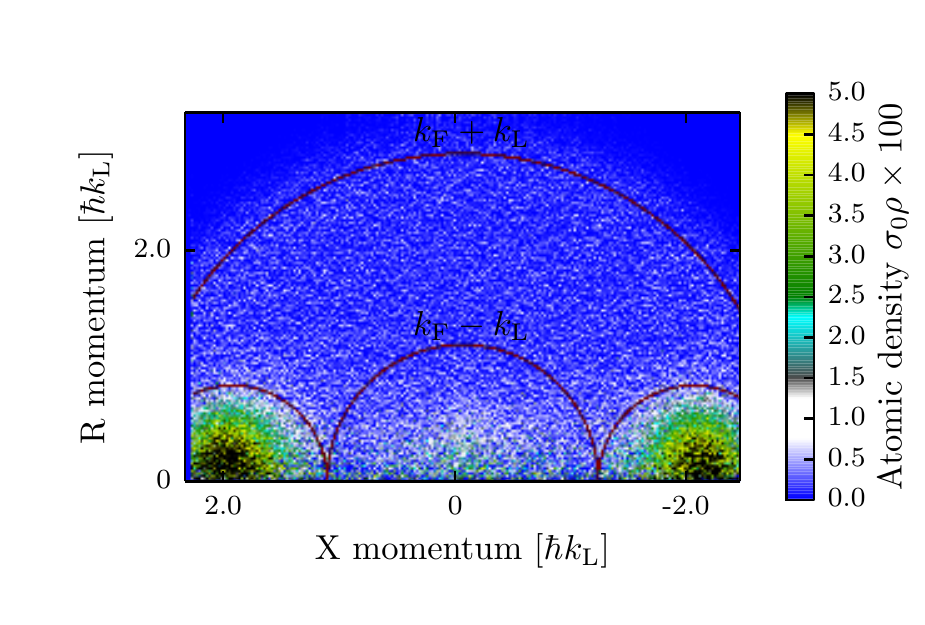}}
\caption{(a) Our experimental setup. Top. The 1D lattice was pulsed, imparting momentum to the atoms and defining the axis of symmetry. Bottom. After time of flight, the two clouds traveling along $\pm$\ex{} directions have separated and the atoms that underwent a single scattering event were evenly distributed in a scattering halo around the unscattered clouds. (b) Inverse Abel transform of corrected image. The atoms within the Fermi momentum $k_F$ of each unscattered cloud center are in the unscattered region and counted towards the total unscattered number. The atoms  within the radius $ k_{\rm{L}}-k_{\rm{F}}\leq r\leq k_{\rm{L}}+k_{\rm{F}}$ but outside the unscattered region are counted towards the number of single scattered atoms.   }
\label{fig:halo}
\end{figure}

We counted the fraction of atoms that experienced a single scattering event for each image. Single scattering events are easily identified, as two atoms that scatter elastically keep the same amplitude of momentum, but depart along an arbitrary direction. Therefore, an atom traveling at $2 \hbar k_{\rm{L}}$ to the right that collides elastically with an atom traveling at $-2 \hbar k_{\rm{L}}$ to the left will depart with equal and opposite momenta $2 \hbar k_{\rm{L}}$ at an arbitrary angle, and after a time-of-flight sufficiently long to convert initial momentum into position, as ours was, such atoms will lie in a spherical shell, producing the scattering halo pictured in Fig. \ref{fig:halo}(a).

Absorption images captured the integrated column density along \ez{}, a projected 2D atomic distribution. To extract the radial dependence of the 3D distribution from the 2D image, we performed a standard inverse Abel transform \cite{FT}. The inverse Abel transform assumes cylindrical symmetry, which was present in our case, with the axis of symmetry along \ex{}, defined by the lattice. We thus obtained the atomic distribution $\rho(r,\theta)$ as a function of $r$, the radial distance from the scattering center, and $\theta$, the angle between $r$ and symmetry axis \ex{}, integrated over $\phi$, the azimuthal angle around the $x$ axis.

We then extracted the number of scattered atoms $N_{\rm{scat}}$ as a fraction of the total atom number $N_{\rm{tot}}$ for each image, as shown in Fig. \ref{fig:halo}(b). The unscattered atom number was the number of atoms in the two unscattered clouds. The number of atoms that underwent a single scattering event was the number of atoms outside the Fermi radius of the unscattered clouds, but inside the arc created by rotating the Fermi momentum $k_{\rm{F}}$ around the original center of the cloud [red arcs in Fig. \ref{fig:halo}(b)]. For both the scattered and unscattered quantities, we extrapolated to include atoms that would fall outside the field of view of our camera. The atoms in the center region were not counted as they were originally in the zero momentum state and could not contribute to the scattering halo under study.

Since we were in the low energy regime (the atomic momentum was much smaller than the momentum set by the van der Waals length $k_{\rm{L}}+k_{\rm{F}}\ll1/l_{\rm{vdW}}$, and we were well below the \pwave{} threshold temperature \cite{DeMarco99}), the scattering cross-section was given by $\sigma=4\pi a^2$. The scattering cross-section $\sigma$ gives the probability $P_{\rm{scat}}=\sigma N/A$ that a single particle will scatter when incident on a cloud of atoms with a surface density of $N/A$, where $A$ is the cross-sectional area of the cloud and $N$ is the number of atoms in the cloud. In our case, each half of the initial cloud, with atoms number $N_{\rm{tot}}/2$, was incident on the other half. Thus, the number of expected scattering events was $N_{\rm{scat}}= (N_{\rm{tot}}/2) \sigma  (N_{\rm{tot}}/2)=\sigma N_{\rm{tot}}^2/4A$. Assuming $A$ was constant for all our data, we defined a fit parameter $b_0=4\pi a_{\rm{bg}}^2/4A$, where $a_{\rm{bg}}$ is the background scattering length. We thus adapted Eq. (\ref{feshbachEq}) to obtain the fit function
\begin{equation}
\frac{N_{\rm{scat}}}{N_{\rm{tot}}^2}=b_0\left(1-\frac{\Delta}{B-B_0}\right)^2 + C,
\label{eq:fit}
\end{equation}
where $B_0$ is the resonant field value and $\Delta$ is the width of the resonance,  the parameters in Eq. (\ref{feshbachEq}), and the offset $C$ accounts for any systematic difference in the initial and final intensity images with no atoms present.

For each value of the bias magnetic field, we took 15 nominally identical images, allowing us to compensate for shot-to-shot atom number fluctuations. We fit the fraction of scattered atoms $N_{\rm{scat}}/N_{\rm{tot}}$ versus the total atom number $N_{\rm{tot}}$ for each of these 15 images to a line. The slope of this fit was taken to be the value of $N_{\rm{scat}}/N_{\rm{tot}}^2$ at that bias magnetic field, and the variance of the fit gave the uncertainty on that data point.

\subsection{Results}
\begin{figure}
	\includegraphics{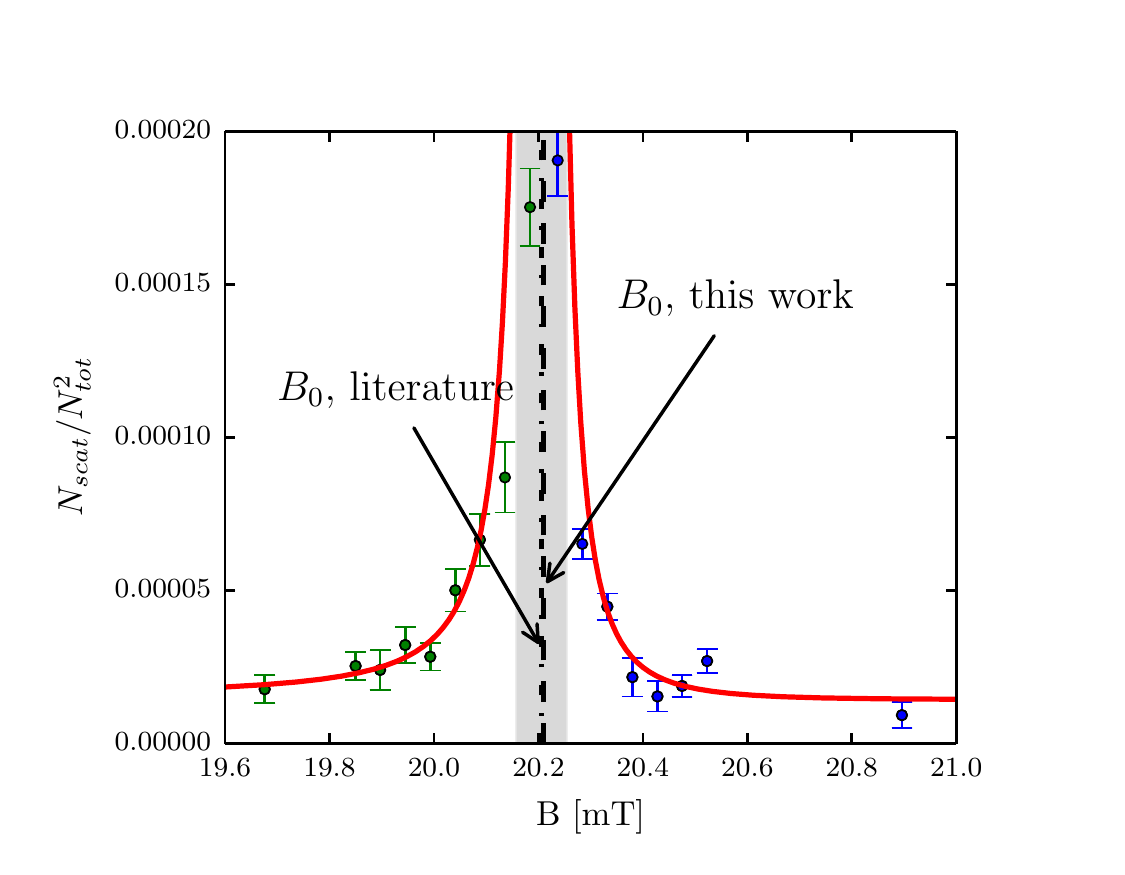}
\caption{Normalized scattered population plotted versus bias field $B$. Green dots represent data taken coming from below the resonance, and blue dots represent the data taken coming from above the resonance. The red curve depicts the line of best fit. The regime where the scattering length is likely large enough for the atoms to behave hydrodynamically is shaded in gray, and data points in that area were excluded from the fit. Resonant field value $B_0$ from literature and as found in this work are indicated. }
\label{fig:fittedFractions}
\end{figure}

Our data is presented in Fig. \ref{fig:fittedFractions}. The red curve depicts a best fit of the model given in Eq. (\ref{eq:fit}). The fit parameters we extracted were $\Delta = 1.0(5)$  \mT{} and $B_0 = 20.206(15)$  \mT{}. To obtain the fit, we used data taken by approaching the resonance from above for points above where we expected the resonance to be and data taken approaching the resonance from below for points below. We also excluded from the fit data points very near the resonance, as there the assumption $\sigma n \ll1$ is no longer valid and the problem must be treated hydrodynamically \cite{Chin10}. Due to this, we could not obtain usable data very close to the resonance, explaining the large uncertainty on the resonance width. 

The accepted values for the $^{40}K$ s-wave Feshbach resonance for the  $\ket{9/2,-9/2}$ and $\ket{9/2,-7/2}$ states are $B_0=20.210(7)$  \mT{} and $\Delta=0.78(6)$  \mT{} \cite{Regal04}, which is in good agreement with our findings. Although the data without the recoil induced detuning correction were $\approx 30\,\%$ different from the corrected data, the optimal parameters from fitting the uncorrected data were within our uncertainties from the values listed above.  Some potential sources of systematic uncertainty that we did not account for include scattering with atoms that did not receive a momentum kick from the lattice pulsing and the impact of multiple scattering events.
\section{Conclusion}
We studied the effects of recoil-induced detuning on absorption images and found an imaging time that maximized SNR to be $\approx40$ \us{} for \K{} atoms. We used these results to directly image s-wave scattering halos of the Fermi gas around the $\approx 20.2$ \mT{} Feshbach resonance and verify the resonance location and width. Our imaging analysis can be used in any absorption imaging application where SNR optimization is critical.
\section*{Acknowledgments}
We thank Marcell Gall for helpful discussions. We thank William R. McGehee and Elizabeth A. Goldschmidt for a careful reading of the manuscript. This work was partially supported by the ARO’s Atomtronics MURI, by the
AFOSR’s Quantum Matter MURI, NIST, and the NSF through the PFC at the JQI. B.K.S. acknowledges support from the National Research Council Research Associateship program. L.M.A. acknowledges support from the National Science Foundation Graduate Research Fellowship program.

\appendix
\section*{Appendix A}
\setcounter{section}{1}
\subsection{Peturbative treatment}
\label{AppendixA1}
 By considering Eqs. (\ref{eq3})-(\ref{eq4}) perturbatively in imaging time, we can obtain corrections to the column density due to recoil induced detuning to second order \cite{LJLthesis}:

\begin{eqnarray}
\label{eq:OD2}
&\sigma_0 n^{\rm{(2)}}  =  c_0+c_1t+c_2t^2, \mbox{ where } \\
 &c_0= \sigma_0 n^{\rm{(1)}}, c_1=0, c_2=\frac{( k_{\rm{r}} v_{\rm{r}})^2}{3}\left[\frac{1}{\tilde{I}_f+1}-\frac{1}{\tilde{I}_0+1}+\mathrm{ln}\left(\frac{\tilde{I}_f+1}{\tilde{I}_0+1}\right)\right].
\end{eqnarray}

\begin{figure}
	\subfigure[]{\includegraphics{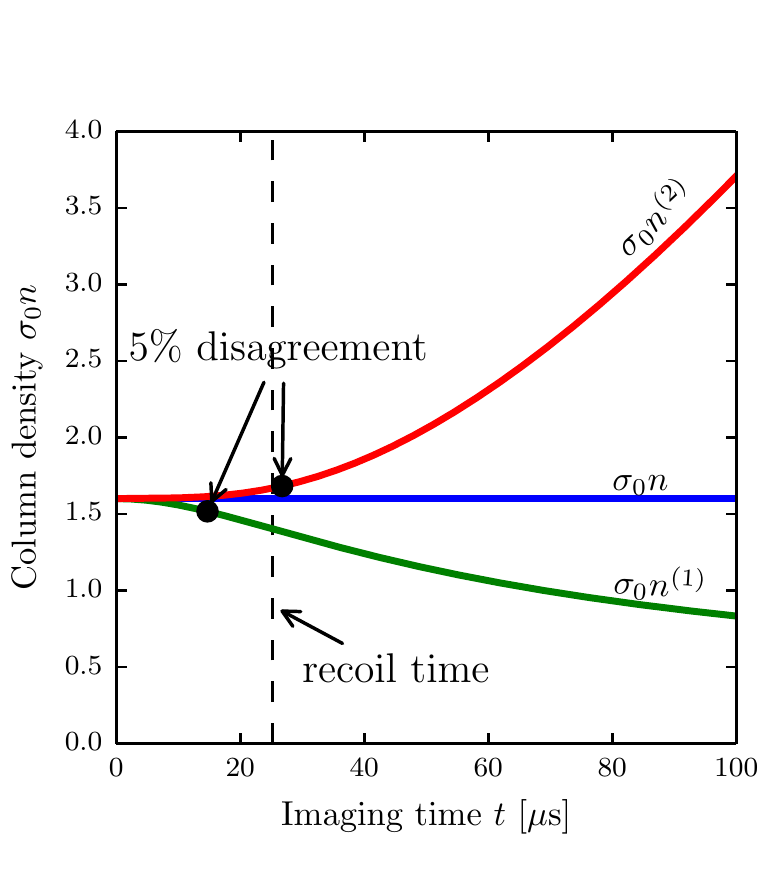}}
	\subfigure[]{\includegraphics{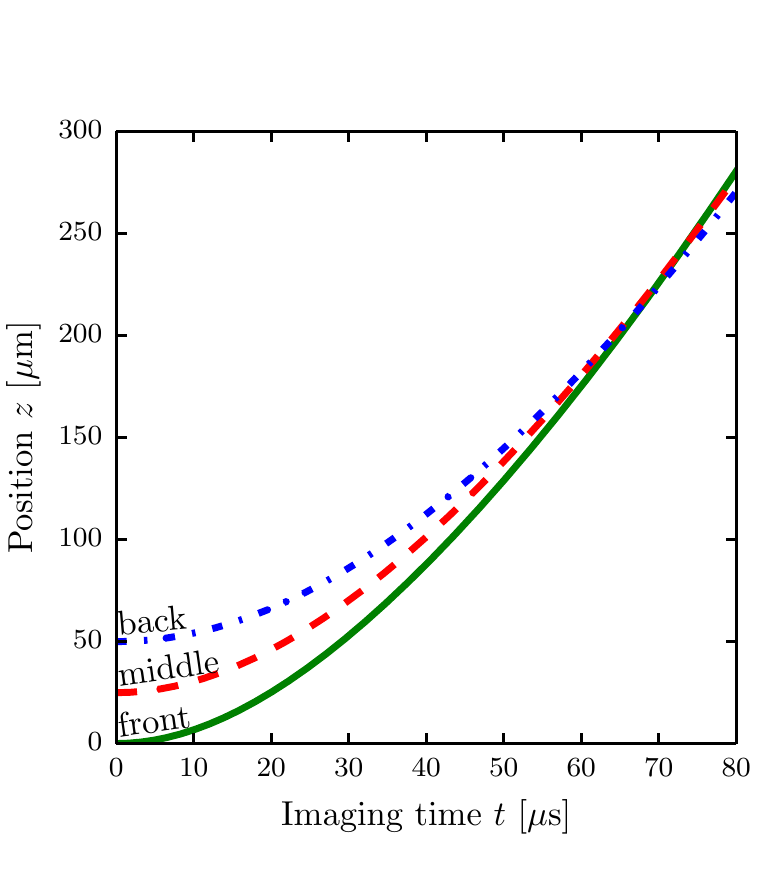}}
\caption{(a) Column densities deduced from simulated optical depths of on-resonant imaging at probe intensity $\tilde{I}_0=0.8$. The input column density was $\sigma_0 n=1.6$. $\sigma_0 n^{\rm{(1)}}$ is the high probe intensity corrected column density given by Eq. (\ref{eq2}). $\sigma_0 n^{\rm{(2)}}$ is the column density as expanded to second order in time, Eq. (\ref{eq:OD2}). (b) Position of atoms as a function of imaging time for atoms in the first (solid green), middle (dashed red), and last (dotted blue) bins of the simulated density distribution for an initial cloud 50 \um{} in extent. The probe intensity used in this calculation was $1.2\, I_{\rm{sat}}$, and the column density was $\sigma_0 n=1.6$.}
\label{fig:simulStuff}
\end{figure}

However, as shown in  Fig. \ref{fig:simulStuff}(a), the perturbative treatment is only accurate to times up to the recoil time $t_{\rm{recoil}}$ \--- defined as the time it takes a single atom to get Doppler shifted from resonance by half a linewidth \--- after which this prediction begins to diverge. To adequately correct for the recoil induced detuning of the atoms, numerical simulation is necessary.

\subsection{Stationary atom model}
\label{AppendixA2}
To solve Eqs. (\ref{eq3})-(\ref{eq4}), we divided the cloud into spatial bins.  In this approximation, the number of atoms in each bin was time-independent.  The algorithm used is shown in Alg. (\ref{algorithm1}), in which we took a Gaussian profile for our initial density distribution. We call the optical depth simulated by this algorithm the simulated optical depth $OD^{\rm{sim1}}$.

\begin{algorithm}
\caption{Stationary atom model}
\label{algorithm1}
\begin{algorithmic}
\STATE $\tilde{I}[n=0,t]=\tilde{I}_0$ \COMMENT{$n$ is the bin index, $t$ is the time index}
\STATE $\tilde{\delta}[n, t=0]=0$ \COMMENT{light initially resonant}
\STATE $H_f=0$ \COMMENT{Radiant fluence seen by camera after passing through cloud}
\FOR[loop over time steps]{$t=0$ to $t_f$}
 \FOR[loop over bins, N is total bin number]{$n=1$ to $N$}
 \STATE $A=\sigma_0\rho[n] dz$ \COMMENT{$dz$ is the size of spatial step}
 \STATE $B=v_{\rm{r}} dt/(\hbar c \rho[n])$  \COMMENT{$dt$ is the size of the time step}
\STATE $\tilde{I}[n,t]=\tilde{I}[n-1,t] - A\tilde{I}[n-1,t]/(1+4\tilde{\delta}[n,t-1]^2+\tilde{I}[n-1,t])$  \COMMENT{Eq. (\ref{eq3})}
\STATE $\tilde{\delta}[n,t]=\tilde{\delta}[n,t-1]+B\left(\tilde{I}[n-1,t]-\tilde{I}[n,t]\right)$  \COMMENT{Eq. (\ref{eq4})}
\ENDFOR
\STATE $H_f =H_f+ \tilde{I}[N,t]dt$ \COMMENT{collecting total fluence seen by the camera}
\ENDFOR
\STATE $OD^{\rm{sim1}}=-\ln{(H_f/\tilde{I}_0t_f)}$
\end{algorithmic}
\end{algorithm}

\par We checked the validity of our simulation in the limits where the problem is analytically solvable. In the limit where the probe intensity is much weaker than the saturation intensity, $\tilde{I}_0\ll 1$, the atoms' velocities are hardly changed, and Eq.(\ref{eq3}) reduces to
\begin{eqnarray}
\frac{d\tilde{I}(z)}{dz}&=-\rho\sigma_0\tilde{I}(z), \mbox{ from which we recover the analytic form }\\
\sigma_0 n^{\rm{(0)}} &= -\ln\tilde{I}_0/\tilde{I}_f. \label{eq6}
\end{eqnarray}
In the limit where the probe intensity is much larger than the saturation intensity, $\tilde{I}_0\gg \tilde{\delta}$, even far detuned atoms will scatter light at their maximum rate. The time dependence of the detuning can thus be neglected, and Eq. (\ref{eq3}) becomes
\begin{eqnarray}
\frac{d\tilde{I}(z)}{dz}&=-\rho\sigma_0, \mbox{ which integrates to }\\
\sigma_0 n &= \tilde{I}_0 - \tilde{I}_f. \label{eq8}
\end{eqnarray}
We recognize the right hand sides of Eq. (\ref{eq6}) and Eq. (\ref{eq8}) as the two terms in Eq. (\ref{eq2}). Thus, as shown in  Fig. \ref{fig:IsatLimits}, $OD^{\rm{sim1}}$  coincides with the optical depth as predicted by Eq. (\ref{eq2}) in both the small and large probe intensity limits.

We used the results of this simulation to check the self-consistency of the  stationary atom assumption, i.e. the distance traveled by the atoms (as deduced from integrating the acquired recoil velocity over the imaging time) is less than the bin size. As shown in Fig. \ref{fig:simulStuff}(b), not only do the atoms travel more than the bin size, but they travel far beyond the initial extent of the cloud. Moreover, owing to the higher initial scatter rate, the back of the cloud overtakes the front for long imaging times. Thus, the atomic distribution as a function of position changes dramatically during the imaging pulse, and the stationary assumption is invalid.
\begin{figure}	
	\subfigure[]{\includegraphics{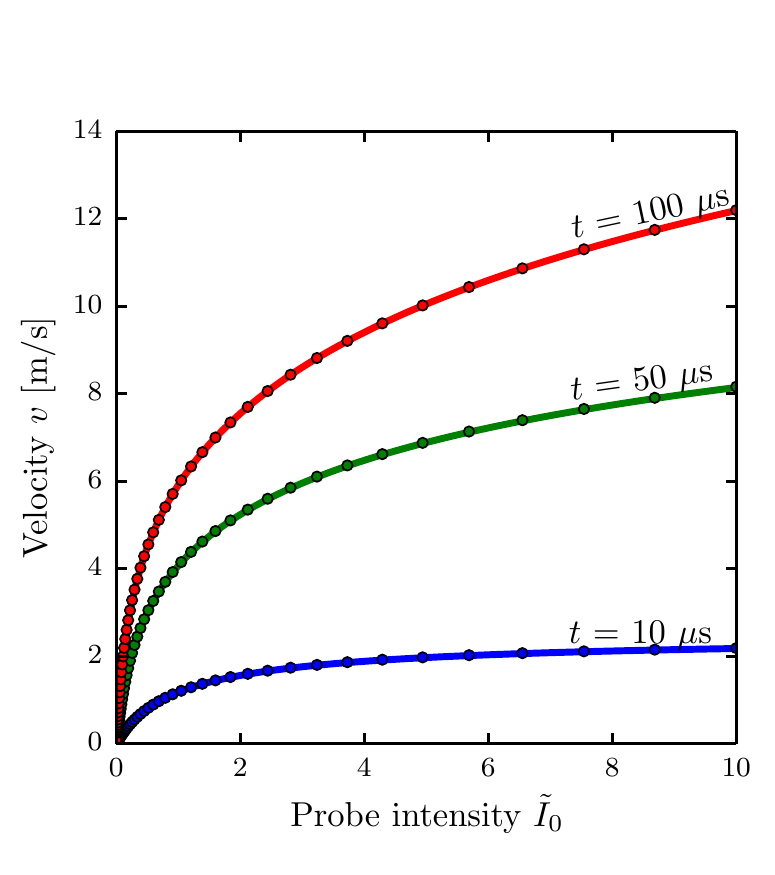}}
	\subfigure[]{\includegraphics{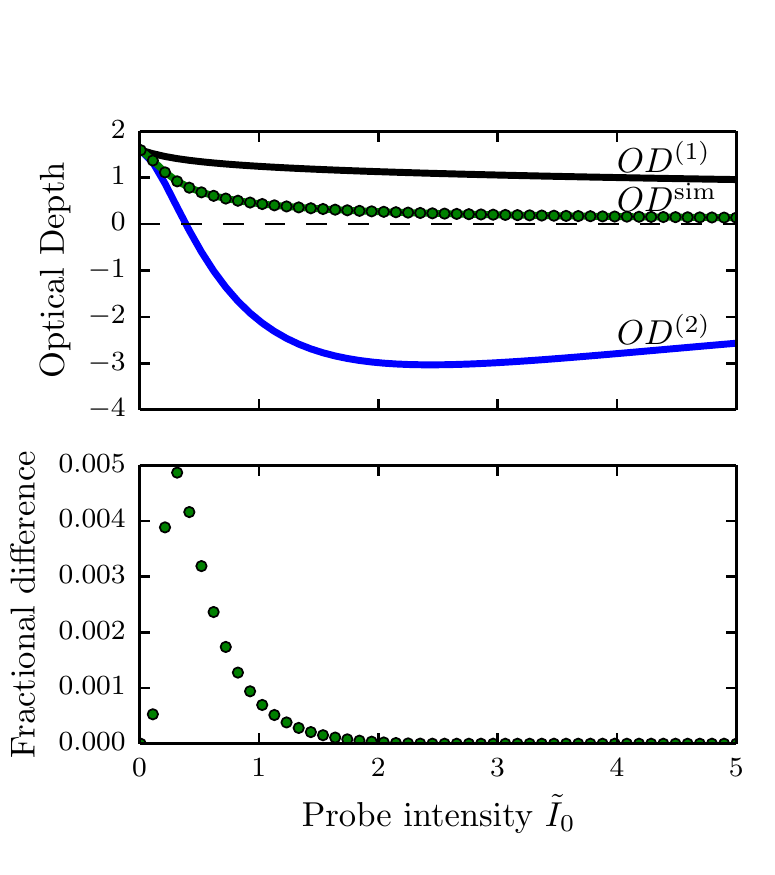}}
\caption{ (a) The velocity of a single composite atom as a function of probe intensity for various imaging times. Simulation data (dots) and numerical  solutions of Eq. (\ref{eq10}) (lines) are in agreement. (b) Top. Optical depth as a function of probe intensity for an imaging time $t=100$ \us. $OD^{\rm{(1)}}$ and $OD^{\rm{(2)}}$ are optical depths predicted from a given column density by Eq. (\ref{eq2}) and (\ref{eq:OD2}) respectively.  The two versions of simulated optical depth, $OD^{\rm{sim1}}$ (green curve) and $OD^{\rm{sim2}}$ (green dots) are plotted. Bottom. The fractional difference between two versions of the simulated $OD$, $\left|OD^{\rm{sim1}}-OD^{\rm{sim2}}\right|/OD^{\rm{sim1}}$.}
\label{fig:simTests}
\end{figure}

\subsection{Traveling atom model}
\label{AppendixA3}
To account for the changing atomic distribution during the imaging pulse, we numerically simulated the classical kinetics of atoms subject to the recoil driven optical forces. To simulate large ensembles in a reasonable time, we modeled composite atoms, each describing the aggregate behavior of $N_{\rm{ca}}$ atoms. The amended algorithm is shown in Alg. (\ref{algorithm2}).
\begin{algorithm}
\caption{Travelling atom model}
\label{algorithm2}
\begin{algorithmic}
\STATE $z[n]=z_0$, $\tilde{\delta}[n]=0$ \COMMENT{initialize position and detuning for each composite atom, labeled by index $n$}
\STATE $O[i]=n$ \COMMENT{make a list of composite atom indexes, ordered by position}
\STATE $\tilde{I}[n=0,t]=\tilde{I}_0$ \COMMENT{$t$ is the time index}
\STATE $H_f=0$ \COMMENT{Radiant fluence seen by camera after passing through cloud}
\FOR[loop over time steps]{$t=0$ to $t_f$}
 \FOR[loop over composite atoms]{$i=1$ to $N_{\rm{ca}}$}
\STATE $n=O[i]$ \COMMENT{apply probe intensity to composite atoms in order of appearance}
 \STATE $A=\sigma_0 N_{\rm{ca}} dz$ \COMMENT{dz is length over which atoms were grouped into single composite atom}
 \STATE $B=v_{\rm{r}} dt/(\hbar c  N_{sa})$  \COMMENT{dt is the time step}
\STATE $\tilde{I}[n,t]=\tilde{I}[n-1,t] - A \tilde{I}[n-1,t]/(1+4\tilde{\delta}[n]^2+\tilde{I}[n-1,t])$  \COMMENT{Eq. (\ref{eq3})}
\STATE $\tilde{\delta}[n]\mathrel{+}=B\left(\tilde{I}[n-1,t]-\tilde{I}[n,t]\right)$  \COMMENT{Eq. (\ref{eq4})}
\STATE $z[n]\mathrel{+}=dt\Gamma\tilde{\delta}/2k_{\rm{r}}$ \COMMENT{$\Gamma\tilde{\delta}/2k_{\rm{r}}$ is the velocity at $\tilde{\delta}$ detuning}
\ENDFOR
\STATE $O[i]$=sort($n$, key =$z[n]$) \COMMENT{sort composite atom indexes by current position}
\STATE $H_f H_f+ \tilde{I}[N,t]dt$ \COMMENT{collecting total fluence seen by the camera}
\ENDFOR
\STATE $OD^{\rm{sim2}}=-\ln{(H_f/\tilde{I}_0t_f)}$
\end{algorithmic}
\end{algorithm}

To validate our code, we again checked the velocity predicted in this model against known limits. One such limit is that of a single composite atom. In this case, there is no attenuation, and the intensity seen by the composite atom is constant at $\tilde{I}_0$. Only the detuning  evolves in time, and Eqs. (\ref{eq3}) and (\ref{eq4}) give
\begin{equation}
\frac{d\tilde{\delta}(t)}{dt}= \frac{ k_{\rm{r}} v_{\rm{r}}}{2} \frac{\tilde{I}}{1+4\tilde{\delta}^2+\tilde{I}}.
\label{eq10}
\end{equation}
Equation (\ref{eq10}) can be solved numerically, and is in agreement with our simulation, as seen in Fig. \ref{fig:simTests}(a).

\begin{figure}
	\subfigure[]{\includegraphics{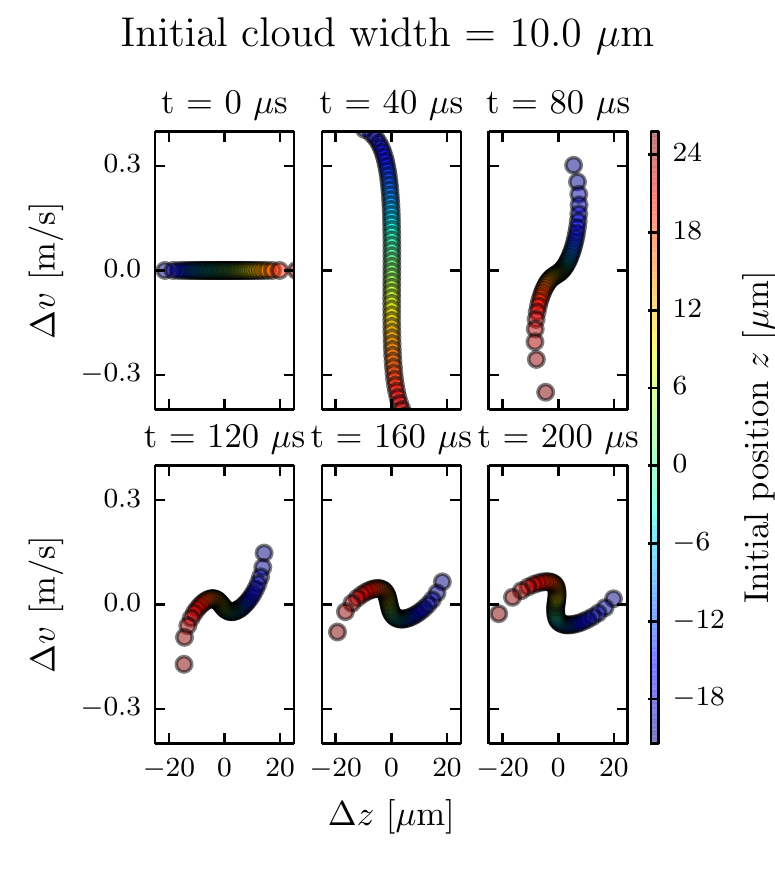}}
	\subfigure[]{\includegraphics{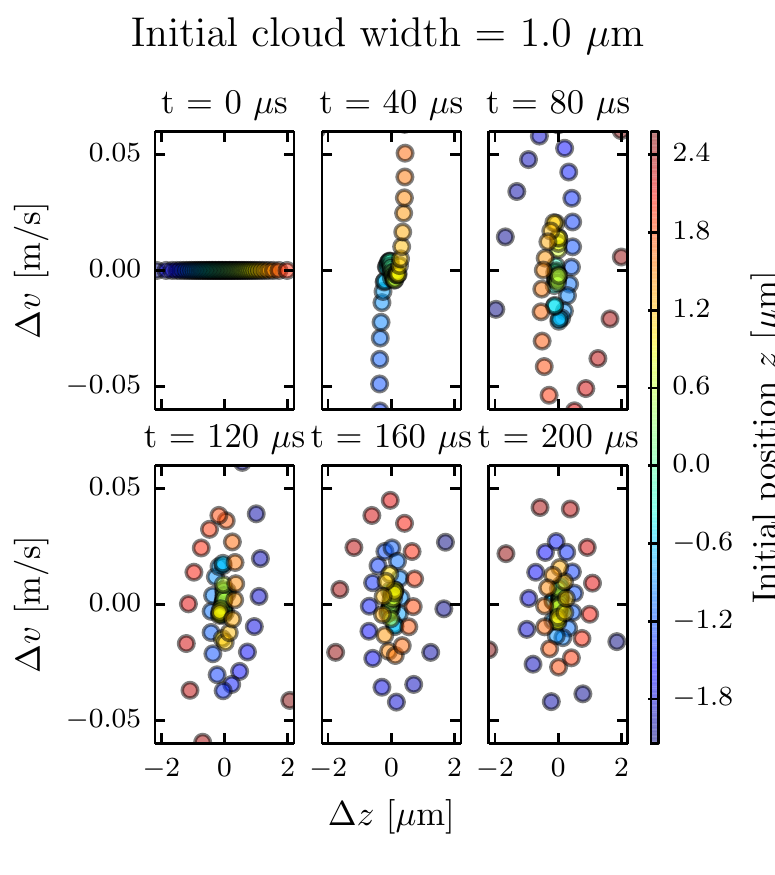}}
\caption{Phase space evolution of an atomic cloud exposed to probe light with intensity $\tilde{I}_0=1.2$. We defined $\Delta v=v -\left< v(t) \right>$  and $\Delta z=z-\left< z(t) \right>$, subtracting out the center of mass position and velocity of the cloud. The column density $\sigma_0 n$ is 1.6, and the initial cloud is a Gaussian with a width of 10 $\mu$m in (a) and 1 $\mu$m in (b). The center of mass velocities $\left< v\right>$ are (0,  3.41, 5.26, 6.52, 7.50, 8.32) m/s sequentially, and are the same for both initial cloud widths. }
\label{fig:phaseSpace}
\end{figure}

We used this model to study the time evolution of the cloud shape during imaging and visualized the phase space evolution of superatoms, shown in Fig. \ref{fig:phaseSpace}. The cloud shape is strongly distorted during imaging.

We compared the optical depths predicted by each of the two models, $OD^{\rm{sim1}}$ and $OD^{\rm{sim2}}$. As seen Fig. \ref{fig:simTests}(b), the predicted optical depths were hardly changed by including the full time evolution:  $\left|OD^{\rm{sim1}}-OD^{\rm{sim2}}\right|/OD^{\rm{sim1}} \le 0.01$ for times up to 100 \us{}, $\tilde{I}_0$ up to 50 and $\sigma_0 n$ up to 2.0. 

\section*{Appendix B}
\label{AppendixB}
We used a Zeeman slower to slow both \Rb{} and \K{} before capturing in a magneto-optical trap (MOT). After 7 s seconds of MOT loading \K{} followed by 1.5 s of loading both \K{} and \Rb{}, we cooled both species in optical molasses for 2 ms. We optically pumped both species into their maximally stretched magnetically trappable states, $\ket{F=9/2, m_F=9/2}$ for \K{} and $\ket{F=2,m_F=2}$ for \Rb{}. Both species were then loaded into a quadrupole magnetic trap with a gradient of $\approx$ 7.68 mT/cm along \ez{}, and cooled evaporatively via forced rf evaporation, sweeping the rf frequency from 18 MHz to 2 MHz in 10 s. The magnetic trap was plugged by a $\lambda =$ 532 nm beam, tightly focused to $\approx$ 30 \um{} and $\approx$ 5 W in power, providing a repulsive potential around the zero field point to prevent Majorana losses. Since the \K{} atoms were spin polarized and therefore only interacted by the strongly suppressed \pwave{} interactions, they re-thermalized largely due to sympathetic cooling with \Rb{} atoms.

We then loaded the atoms into a crossed optical dipole trap, provided by a 1064 nm fiber laser, and continued evaporative cooling by slowly ramping down the dipole trap to trap frequencies of $(\omega_x,\omega_y,\omega_z)/2\pi =(39, 42, 124)$ Hz (for potassium atoms) in the three spatial directions, while also turning off the quadrupole field. We then used adiabatic rapid passage (ARP) to transfer the \Rb{} atoms from the $\ket{F=2, m_F=2}$ state to the  $\ket{F=1, m_F=+1}$ absolute ground state via 6.8556 GHz microwave coupling (20.02 MHz from the zero field resonance) followed by a magnetic field sweep from -0.469 mT to -0.486 mT in 50 ms. This state was chosen to minimize spin changing collisions with \K{} atoms during any further evaporation \cite{BestThesis}.  We then briefly applied an on-resonant probe laser, ejecting any remaining \Rb{} atoms in the $F=2$ manifold from the trap. We again used ARP to transfer the \K{} atoms into the $\ket{F=9/2,m_F=-9/2}$ state by using a 3.3 MHz rf field and sweeping the bias magnetic field from -0.518 mT to -0.601 mT in 150 ms.

\section*{References}
\bibliography{refs}{}

\begin{thebibliography}{10}

\bibitem{Greiner03}
M.~Greiner, C.~A. Regal, and D.~S. Jin.
\newblock {Emergence of a molecular Bose-Einstein condensate from a Fermi gas}.
\newblock {\em Nature}, 426(6966):537--540, 2003.

\bibitem{Zwierlein03}
M.~W. Zwierlein, C.~A. Stan, C.~H. Schunck, S.~M.~F. Raupach, S.~Gupta,
  Z.~Hadzibabic, and W.~Ketterle.
\newblock {Observation of Bose-Einstein condensation of molecules}.
\newblock {\em Phys. Rev. Lett.}, 91:250401, 2003.

\bibitem{Jochim03}
S.~Jochim, M.~Bartenstein, A.~Altmeyer, G.~Hendl, S.~Riedl, C.~Chin,
  J.~Hecker~Denschlag, and R.~Grimm.
\newblock {Bose-Einstein condensation of molecules}.
\newblock {\em Science}, 302(5653):2101--2103, 2003.

\bibitem{Bartenstein04}
M.~Bartenstein, A.~Altmeyer, S.~Riedl, S.~Jochim, C.~Chin, J.~Hecker Denschlag,
  and R.~Grimm.
\newblock {Collective excitations of a degenerate gas at the BEC-BCS
  crossover}.
\newblock {\em Phys. Rev. Lett.}, 92:203201, 2004.

\bibitem{Bourdel04}
T.~Bourdel, L.~Khaykovich, J.~Cubizolles, J.~Zhang, F.~Chevy, M.~Teichmann,
  L.~Tarruell, S.~J. J. M.~F. Kokkelmans, and C.~Salomon.
\newblock {Experimental study of the BEC-BCS crossover region in Lithium 6}.
\newblock {\em Phys. Rev. Lett.}, 93:050401, 2004.

\bibitem{Zwierlein04}
M.~W. Zwierlein, C.~A. Stan, C.~H. Schunck, S.~M.~F. Raupach, A.~J. Kerman, and
  W.~Ketterle.
\newblock {Condensation of pairs of Fermionic atoms near a Feshbach resonance}.
\newblock {\em Phys. Rev. Lett.}, 92:120403, 2004.

\bibitem{Regal04}
C.~A. Regal, M.~Greiner, and D.~S. Jin.
\newblock {Observation of resonance condensation of fermionic atom pairs}.
\newblock {\em Phys. Rev. Lett.}, {92}({4}), {2004}.

\bibitem{Chin10}
C.~Chin, R.~Grimm, P.~Julienne, and E.~Tiesinga.
\newblock {Feshbach resonances in ultracold gases}.
\newblock {\em Rev. Mod. Phys.}, 82:1225--1286, 2010.

\bibitem{Timmermans99}
E.~Timmermans, P.~Tommasini, M.~S. Hussein, and A.~Kerman.
\newblock {Feshbach resonances in atomic Bose-Einstein condensates}.
\newblock {\em {Physics Reports-review Section Of Physics Letters}},
  {315}({1-3}):{199--230}, {1999}.

\bibitem{Tiesinga93}
E.~Tiesinga, B.~J. Verhaar, and H.~T.~C. Stoof.
\newblock {Threshold and resonance phenomena in ultracold ground-state
  collisions}.
\newblock {\em Phys. Rev. A}, {47}({5, B}):{4114--4122}, {1993}.

\bibitem{Lysebo09}
M.~Lysebo and L.~Veseth.
\newblock {\textit{Ab initio} calculation of Feshbach resonances in cold atomic
  collisions: $s$- and $p$-wave Feshbach resonances in $^{6}\rm{Li}_{2}$}.
\newblock {\em Phys. Rev. A}, 79:062704, 2009.

\bibitem{Gao11}
B.~Gao.
\newblock {Analytic description of atomic interaction at ultracold
  temperatures. II. Scattering around a magnetic Feshbach resonance}.
\newblock {\em Phys. Rev. A}, 84:022706, 2011.

\bibitem{Inouye98}
S.~Inouye, M.~R. Andrews, J.~Stenger, H.~J. Miesner, D.~M. Stamper-Kurn, and
  W.~Ketterle.
\newblock {Observation of Feshbach resonances in a Bose-Einstein condensate}.
\newblock {\em {Nature}}, {392}({6672}):{151--154}, {1998}.

\bibitem{Cornish00}
S.~L. Cornish, N.~R. Claussen, J.~L. Roberts, E.~A. Cornell, and C.~E. Wieman.
\newblock {Stable Rb-85 Bose-Einstein condensates with widely tunable
  interactions}.
\newblock {\em Phys. Rev. Lett.}, {85}({9}):{1795--1798}, {2000}.

\bibitem{Regal03}
C.~A. Regal and D.~S. Jin.
\newblock {Measurement of positive and negative scattering lengths in a Fermi
  gas of atoms}.
\newblock {\em Phys. Rev. Lett.}, 90:230404, 2003.

\bibitem{OHara02}
K.~M. O'Hara, S.~L. Hemmer, M.~E. Gehm, S.~R. Granade, and J.~E. Thomas.
\newblock {Observation of a strongly interacting degenerate Fermi gas of
  atoms}.
\newblock {\em Science}, 298(5601):2179--2182, 2002.

\bibitem{Monroe93}
C.~Monroe, E.~Cornell, C.~Sackett, C.~Myatt, and C.~Wieman.
\newblock {Measurement of Cs-Cs elastic scattering at \textit{T} =30 $\mu${}K}.
\newblock {\em Phys. Rev. Lett.}, 70:414--417, 1993.

\bibitem{Chikkatur00}
A.~P. Chikkatur, A.~G\"orlitz, D.~M. Stamper-Kurn, S.~Inouye, S.~Gupta, and
  W.~Ketterle.
\newblock {Suppression and enhancement of impurity scattering in a
  Bose-Einstein condensate}.
\newblock {\em Phys. Rev. Lett.}, 85:483--486, 2000.

\bibitem{Wilson04}
Nicholas~R. Thomas, Niels Kj\ae{}rgaard, Paul~S. Julienne, and Andrew~C.
  Wilson.
\newblock {Imaging of $s$ and $d$ partial-Wave interference in quantum
  scattering of identical Bosonic atoms}.
\newblock {\em Phys. Rev. Lett.}, 93:173201, Oct 2004.

\bibitem{Williams2012}
R.~A. Williams, L.~J. LeBlanc, K.~Jiménez-García, M.~C. Beeler, A.~R. Perry,
  W.~D. Phillips, and I.~B. Spielman.
\newblock {Synthetic partial waves in ultracold atomic collisions}.
\newblock {\em Science}, 335(6066):314--317, 2012.

\bibitem{Aikawa14}
K.~Aikawa, A.~Frisch, M.~Mark, S.~Baier, R.~Grimm, and F.~Ferlaino.
\newblock {Reaching Fermi degeneracy via universal dipolar scattering}.
\newblock {\em Phys. Rev. Lett.}, 112:010404, Jan 2014.

\bibitem{Lu12}
M.~Lu, N.~Q. Burdick, and B.~L. Lev.
\newblock {Quantum degenerate dipolar Fermi gas}.
\newblock {\em Phys. Rev. Lett.}, 108:215301, May 2012.

\bibitem{LCT}
H.J. Metcalf and P.~van~der Straten.
\newblock {\em Laser Cooling and Trapping}.
\newblock Graduate Texts in Contemporary Physics. Springer New York, 1999.

\bibitem{Reinaudi07}
G.~Reinaudi, T.~Lahaye, Z.~Wang, and D.~Gu\'{e}ry-Odelin.
\newblock {Strong saturation absorption imaging of dense clouds of ultracold
  atoms}.
\newblock {\em Opt. Lett.}, 32(21):3143--3145, 2007.

\bibitem{Konstantinidis12}
G.~O. Konstantinidis, M.~Pappa, G.~Wikstroem, P.~C. Condylis, D.~Sahagun,
  M.~Baker, O.~Morizot, and W.~von Klitzing.
\newblock {Atom number calibration in absorption imaging at very small atom
  numbers}.
\newblock {\em {Central European Journal Of Physics}}, {10}({5}):{1054--1058},
  {2012}.

\bibitem{Tiecke}
Tobias~G. Tiecke.
\newblock {\em {Feshbach resonances in ultracold mixtures of the fermionic
  quantum gases $^6\rm{Li}$ and $^{40}\rm{K}$.}}
\newblock PhD thesis, University of Amsterdam, 2009.

\bibitem{Williams13}
R.~A. Williams, M.~C. Beeler, L.~J. LeBlanc, K.~Jim\'enez-Garc\'{i}a, and I.~B.
  Spielman.
\newblock {Raman-induced interactions in a single-component Fermi gas near an
  $s$-wave Feshbach resonance}.
\newblock {\em Phys. Rev. Lett.}, 111:095301, 2013.

\bibitem{Lin09}
Y.-J. Lin, A.~R. Perry, R.~L. Compton, I.~B. Spielman, and J.~V. Porto.
\newblock {Rapid production of $^{87}\rm{Rb}$ Bose-Einstein condensates in a
  combined magnetic and optical potential}.
\newblock {\em Phys. Rev. A}, 79:063631, 2009.

\bibitem{KarinaThesis}
Karina Jimenez-Garcia.
\newblock {\em {Artificial gauge fields for ultracold neutral atoms}}.
\newblock PhD thesis, Joint Quantum Institute, National Institute of Standards
  and Technology, and the University of Maryland, 2012.

\bibitem{DeMarco99}
B.~DeMarco, J.~L. Bohn, J.~P. Burke, M.~Holland, and D.~S. Jin.
\newblock {Measurement of $\mathit{p}$-wave threshold law using evaporatively
  cooled Fermionic atoms}.
\newblock {\em Phys. Rev. Lett.}, 82:4208--4211, 1999.

\bibitem{Wu05}
S.~Wu, Y.-J. Wang, Q.~Diot, and M.~Prentiss.
\newblock {Splitting matter waves using an optimized standing-wave light-pulse
  sequence}.
\newblock {\em Phys. Rev. A}, 71:043602, 2005.

\bibitem{Edwards10}
M.~Edwards, B.~Benton, J.~Heward, and C.~W. Clark.
\newblock Momentum-space engineering of gaseous bose-einstein condensates.
\newblock {\em Phys. Rev. A}, 82:063613, 2010.

\bibitem{FT}
R.~Bracewell.
\newblock {\em The Fourier Transform and its Applications}.
\newblock McGraw-Hill, New York, 1965.

\bibitem{LJLthesis}
Lindsey~J. LeBlanc.
\newblock {\em {Exploring many-body physics with ultracold atoms}}.
\newblock PhD thesis, University of Toronto, 2011.

\bibitem{BestThesis}
Thorseten Best.
\newblock {\em Interacting Bose-Fermi mixtures in optical lattices}.
\newblock PhD thesis, Johannes Gutenberg-Universitat, 2010.

\end{thebibliography}
\bibliographystyle{unsrt}

\end{document}